\documentclass[preprint]{revtex4}
\usepackage{graphicx}
\def\be{\begin{equation}}
\def\ee{\end{equation}}
\def\ba{\begin{eqnarray}}
\def\ea{\end{eqnarray}}

\begin{document}
\parskip 3pt
\renewcommand{\topfraction}{1.2}
\vskip 3cm
\title {\Large\bf Classical Propagation of Strings across a
  Big Crunch/Big Bang Singularity} 
\author{\bf Gustavo Niz\footnote{\tt G.Q.Niz@damtp.cam.ac.uk} and Neil
  Turok\footnote{\tt N.G.Turok@damtp.cam.ac.uk}} 
\affiliation{ {DAMTP, Centre for Mathematical Sciences, 
Wilberforce Rd, Cambridge, CB3
    0WA, UK.}}
 {\begin{abstract} 
One of the simplest time-dependent solutions of M theory
consists of nine-dimensional Euclidean space 
times 1+1-dimensional compactified Milne space-time.
With a further modding out by $Z_2$, the space-time represents two 
orbifold planes which collide and re-emerge, a process proposed as 
an explanation of the hot big bang~\cite{ekpyrotic,cyclic,turok}. 
When the two planes are near, the light
states of the theory consist of winding
M2-branes, describing fundamental strings in a particular
ten-dimensional background.
They suffer no blue-shift as the M theory dimension
collapses, and 
their equations of motion are regular
across the transition from big crunch to big bang.
In this paper, we study the
classical
evolution of fundamental strings across the 
singularity in some detail. We also develop a 
simple semi-classical approximation to the quantum 
evolution which
allows one to compute 
the quantum production of excitations on the string and 
implement it in a simplified example.
\end{abstract}}
\maketitle

\section{Introduction}

Recently, an M theory model of a 
big crunch/big bang transition has been proposed.
The model consists of two empty, parallel orbifold planes
colliding and re-emerging.
Close to the collision, the spectrum of the theory 
splits into two types of states. The light states are
winding M2-branes describing perturbative 
string theory, including gravity. The massive 
states, corresponding to higher Kaluza-Klein modes associated
with the M theory dimension, are non-perturbative 
D0-brane states whose masses diverge
as the collision approaches. 
In Ref.~\cite{turok}, it was shown that the equations of motion
of the winding M2-branes
are regular 
at the brane collision and yield an 
an unambiguous classical evolution across it. Since these states 
describe gravity in quantized string theory, the suggestion is that gravity 
is better-behaved in this situation
than it is in general relativity. 
If this is true then, rather than seeding
a gravitational instability, the massive Kaluza-Klein states 
 may simply decouple around the collision. 
Provided their density is negligible in the incoming state,
as it is in the ekpyrotic and cyclic 
models~\cite{ekpyrotic,cyclic,wesley},
then the entire transition may be 
describeable using the perturbative string theory states alone.

In this paper, we will assume the massive states can
be neglected and focus on the evolution of the fundamental
string states across a big crunch/big bang transition of
this type.
By solving the string evolution equations numerically 
we show how the higher string modes 
become excited in a well-defined way as strings cross 
the transition. We also develop a simple approximation
whereby quantum production of excited modes on the string
may be computed in a semi-classical manner. Related work
may be found in Ref.~\cite{tolley}. Earlier work on
strings in cosmological backgrounds is reviewed in 
Ref.~\cite{sanchez}.

The background space-time we are interested in 
is most simply understood
in eleven-dimensional terms.
Near the orbifold plane collision, the 
line element reduces to that for 
a $Z_2$-compactified Milne universe times $R^9$, namely
\be\label{milne11d}
ds^2=-dt^2+t^2d\theta^2+\sum_{i=1}^9(dx^i)^2,
\ee
where $0\leq \theta\leq \theta_0$, and a $Z_2$ reflection is
imposed about each end point. 
For $t<0$ ($t>0$), the two orbifold planes approach (recede)
with relative rapidity $\theta_0$. Away from the
singularity at 
$t=0$ the space-time is flat and hence an
automatic solution of any theory governed by field equations
involving purely geometrical terms
~\cite{geom}.
At $t=0$ the metric degenerates and the 
equations of general relativity become singular.
Nevertheless, one can 
analytically continue the background solution
through $t=0$. As mentioned, the equations
governing the winding M2-branes are 
regular at $t=0$, so the M2-branes evolve smoothly
across the big crunch/big bang transition.

The winding M2-branes reduce to fundamental
strings when the M theory dimension is small,
so one can also describe the situation in ten-dimensional
terms. In string frame, the dimensionally-reduced metric
takes the form
$g_{\mu \nu}= |t/t_s| \eta_{\mu \nu}$, with a time-dependent
dilaton, 
$\phi= {3\over 2} {\rm ln} |t/t_s|$. Here $e^\phi$ is the
string coupling and 
$t_s$ is by definition the time when the string coupling
is unity
~\cite{seiberg}. For $|t|\ll t_s$, stringy interactions
are weak and one can, to a first approximation, 
treat the strings as free. For 
$|t|$ greater than or of order $t_s$, however, the string
theory is at strong coupling and one must switch to an
eleven-dimensional supergravity description 
~\cite{witten,ovrut}. Recently, techniques
for solving the relevant higher-dimensional Einstein
equations have been developed~\cite{paul}, which
can be applied to this situation. As long as 
one is considering long-wavelength cosmological
perturbations, evolving classically, the transition
from the supergravity regime to the string theory
($\alpha'$ expansion) regime appears unproblematic: 
the eleven-dimensional
Einstein equations should reduce to the appropriate
ten dimensional Einstein-dilaton effective theory as
the branes become near.
In this paper, our main focus is on the classical dynamics
of the string near $t=0$.
In a companion paper we provide a complementary treatment,
studying the stringy
$\beta$ function equations in the usual
$\alpha'$ expansion to see whether 
higher order corrections 
are significant when the string scale 
crosses the Hubble radius~\cite{alphacorrections}.

In the eleven dimensional picture, there is no dilaton field.
The unique mass scale is set by the M2-brane tension 
$\mu_2 \equiv M_{11}^3$, where
$M_{11}$ is
the eleven-dimensional Planck mass
~\cite{tension}.
Winding membranes, of
length $\theta_0 |t|$ behave like strings with a time-dependent
tension, $\mu_1= M_{11}^3 \theta_0 |t|$, in Minkowski space-time. 
One can equally well view the strings as having a
fixed tension, $\mu_1= M_{11}^3 \theta_0 |t_s|$, but living
in a string-frame background metric, $g_{\mu \nu}^s= |t/t_s|\eta_{\mu \nu}$.
The string coupling constant 
is set by the size of the M theory dimension in eleven-dimensional
Planck units: $e^\phi =(M_{11} \theta_0 |t|)^{3\over 2}\equiv |t/t_s|^{3\over 2}$.

Starting well away from the collision, at 
large times $|t|$ the system is described by
eleven-dimensional supergravity. As one approaches
the collision the 
string coupling falls below unity, when $|t|$ falls below 
\be
t_s= \theta_0^{-1} M_{11}^{-1}.
\label{stw}
\ee
When $|t|=t_s$, the string mass scale
is $M_{11}$ and the string length 
is $l_s \sim M_{11}^{-1}
\sim \theta_0 t_s$. 
For small $\theta_0$, $l_s$ is far smaller than the 
Hubble horizon scale $t_s$ so the 
characteristic oscillations of the string are 
little affected by the expansion of the universe. 
Corrections due to the 
background space-time curvature are small and
the usual $\alpha'$ expansion holds good. However, as $|t|$ decreases
further,
the physical size of an oscillating string remains 
approximately constant while the physical Hubble radius 
falls as $|t| (|t|/t_s)^{1\over 2}$. 
The string length
crosses the Hubble radius at a time 
\be
t_X \sim \theta_0^{2\over 3} t_s = \theta_0^{-{1\over 3} } M_{11}^{-1}.
\label{stx}
\ee
Thereafter, the string tends towards an ultralocal evolution in which 
the curvature of the string is unimportant and each bit of
the string evolves independently. This is the regime of the 
$1/\alpha'$ expansion discussed in Ref.~\cite{turok}.
Figure 1 illustrates the three regimes: from strong to weak
coupling at $t_s \sim \theta_0^{-1} M_{11}^{-1}$
and from flat space to ultralocal evolution
at $t_X \sim \theta_0^{-{1\over 3}}M_{11}^{-1}$. For small $\theta_0$, as
we shall assume throughout this paper, these
times are well-separated. 

\begin{figure}[t!]
{\centering
\resizebox*{6in}{2.3in}{\includegraphics{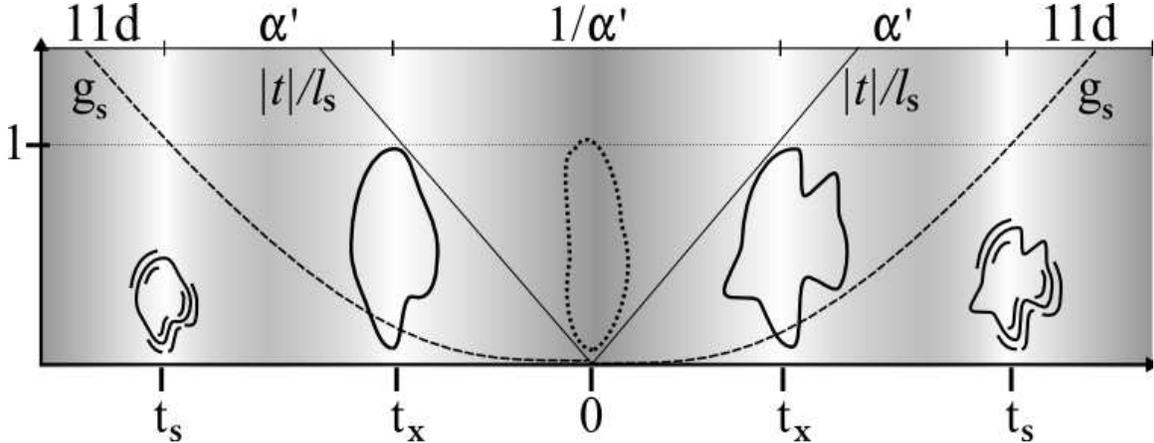}} }
\caption{Three phases of an orbifold plane collision, with
rapidity $\theta_0<1$. Cosmological
perturbations are created when the effective theory is the low energy
limit of M-theory in eleven dimensions. After the orbifolds get
closer and the string coupling ($g_s\sim |t/t_s|^{3/2}$) falls below
unity, the system is better described 
in terms of perturbative string theory and the usual expansion
in $\alpha'$. A second transition
occurs when the string scale leaves the Hubble
radius. Here the string tension becomes dynamically subdominant
and the theory is better described by an expansion in 
$1/\alpha'$. In this regime the string is better-described as 
a collection of independent `bits' whose evolution across
the singularity is smooth. After $t=0$ the string tension is
reconstituted and the system runs back through the two
abovementioned transitions. 
}
\label{diagram}
\end{figure}

In this paper, we wish to study the weak coupling regime,
$|t|<t_s$, during which the strings  
cross the singularity. In this regime, for $|t|<t_X$,
the usual $\alpha'$ expansion fails and we must replace it
with an expansion in the string tension, proportional
to $1/\alpha'$. 
Our ultimate goal is to compute a `mini S-matrix',
evolving weakly-coupled incoming string states
from an initial time just after $-t_s$ to a final
time just before $+t_s$. 
In this paper we shall make modest progress towards this goal
by computing the 
`mini S-matrix' 
in a first, classical approximation.

Any classical solution of the string equations also solves the 
Heisenberg operator equations for the string field $x^\mu(\tau,\sigma)$, 
to leading order in $\hbar$. For highly excited string states,
involving large occupation numbers, the classical approximation
should be reasonable. However, to describe gravity we
are really more interested in the low-lying states. 
For these states, quantum corrections are likely to
be significant and there is less we can say with precision. 

Nevertheless, there are arguments which suggest that
quantum corrections may be manageable in the relevant regime.
The usual $\alpha'$ expansion is based on the approximating
space-time as locally flat, and solving the string equations
of motion in an expansion in the space-time curvature. This
approximation clearly fails to describe even the classical
motion of the fundamental strings near $t=0$. Hence the
failure of the $\alpha'$ expansion in the quantum theory 
does not, on its own, 
imply that quantum
corrections to the string evolution are large\cite{devega}. Second, 
the key phenomena relevant to the small time regime take
place when the background curvature length ({\it i.e.} the Hubble 
radius)
falls below the string scale. The string dynamics 
becomes ultralocal in this regime: in 
cosmological terms, the string is `super-horizon' and hence
may be expected to evolve in a classical manner. 

In flat spacetime, the most dramatic consequence of
quantizing the string is that the first excited
states of the string are massless instead of massive 
as a classical treatment would suggest. This is
due to quantum mechanical 
renormalization of the mass squared operator, 
$m^2 = (p^0)^2-\vec{p}^2$, where $p^\mu$ is the string's
centre of mass momentum. The renormalized 
mass squared 
operator is lowered by one quantized unit relative to the
normal ordered classical expression, and this lowering
causes the first excited states 
to become massless. As a result, the string centre of mass
trajectory $x_{cm}^\mu = q^\mu + p^\mu \tau$ is altered
from the classical expression 
by $(p^0)^2 \rightarrow (p^0)^2 - M^2$ with $M$ a constant
so that the quantum corrected trajectory is null instead
of spacelike. Note, however, that this 
shift in the energy $p^0$ merely alters the 
time coordinate of the centre of mass of the string and it
does not affect the string spatial 
coordinates and momenta, which obey exactly
the same equations as in the classical theory.
Again, this discussion suggests that quantum
effects may be relatively modest in the situation
we are interested in.

In this paper, we focus on the classical 
evolution of the string across $t=0$. 
Starting at a time $t$ of order $-t_s$, we expect the string to
oscillate 
nearly adiabatically and
with almost fixed physical size until the string scale 
crosses the 
Hubble radius at time $t=-t_X$. For small $\theta_0$, 
$t_s\gg  
t_X$, hence there is a large range of time
over which adiabatic evolution holds. This allows us 
to clearly identify the incoming and outgoing states in
terms of the 
usual flat space-time modes. 
As we have already mentioned, the natural size scale for the
incoming states we are interested in (i.e. the graviton, 
dilaton or antisymmetric tensor states) 
is the string scale at $t_s$,
\be
\Delta x\sim \ M_{11}^{-1} \sim \theta_0 t_s.
\ee
By the same token, the natural measure for 
the spatial momentum of the string state at this time is
\be
P \sim  M_{11}\sim (\theta_0 t_s)^{-1},
\ee
and we shall always measure the momentum of the incoming string in
this basic units. 

One of the nice features of the problem at hand is that 
the string tends to simple flat space evolution at 
large asymptotic times:
\be
x^\mu(\tau, \sigma) \rightarrow\, ^{(in)}\!\,\,\!x^\mu(\tau,\sigma) \qquad t \rightarrow - t_s,
\ee
and 
\be
x^\mu(\tau,\sigma) \rightarrow \, ^{(out)}\!\,\,\!x^\mu(\tau,\sigma) \qquad t \rightarrow +t_s,
\ee
where $^{in}x^\mu(\tau,\sigma)$ and $^{out}x^\mu(\tau,\sigma)$ are free string
fields evolving adiabatically in the usual flat space string modes.
The limits should, strictly speaking, be expressed in terms of a 
$t_c <t_s$ so that we are always
studying times for which the string coupling is smaller than unity.
In practice this will not be an important distinction since 
as long as $\theta_0$ is small, the string states we study
tend to their asymptotic behavior at times $t_X \ll t_s$.

In a linear field theory in a time-dependent background, 
where the field tends to the usual Minkowski evolution 
in the asymptotic past, 
{\it all} of the information regarding 
quantum amplitudes may be obtained from real solutions
of the classical field equations (see
e.g. ~\cite{birrell}). One simply expresses the incoming 
field as a linear combination of creation and annihilation operators,
defined by their action on the incoming vacuum state, multiplied by
the appropriate positive and negative frequency modes. 
The classical field equations then determine the
quantum Heisenberg field for all time. Using the time-dependent
field, one can then 
compute any correlation function of interest,
evaluated in any incoming state expressed in terms of
incoming creation operators acting on the incoming vacuum. 

The main difference in our situation is that the string 
evolution is {\it nonlinear}. Nevertheless, the nonlinearity
turns out to be of a very simple form, where one can apply a
semi-classical approximation in a self-consistent way.
One can follow a similar procedure to that in the linear case
to compute the quantum production of string excitations
due to passage across the singularity. We shall explain
this calculation, for a specially simple case -- a circular 
loop -- in the final section of this paper. 
The calculation may
be generalized to higher modes
of the string and we shall do so in a future publication.

The outline of this paper is as follows. 
In section II we present the equation governing 
winding membranes, discuss their adiabatic solutions
and define various quantities of interest. In section III
we study the classical evolution of a circular loop, 
representing the dilaton and its massive counterparts.
In section IV we study the classical evolution of 
a rotor, representing the modes of maximal
spin (for example, the graviton), at each mass level.
Section V describes the classical evolution of a 
state of intermediate angular momentum, representing an
antisymmetric tensor state and massive analogs. 
Section VI gives an
example of a quantum transmutation amplitude 
computed for an incoming dilaton-like state. 
In section VII we give a brief summary
with future proposals. In Appendix A we
give an analytic solution method for the classical
string equations, representing an expansion in powers of
the string tension, {\it i.e.}, inverse powers of
$\alpha'$, which accurately describes the passage
of a circular loop across
the singularity. Intriguingly, in this 
regime polylogarithm functions enter, perhaps hinting 
at some deeper analytic solution yet to be found. We also present
an analytic solution to second order in $1/\alpha'$ for 
a more general string configuration. Finally, Appendix B is devoted to
some numerical checks on the right/left mover decomposition 
for the special case of a classical rotor.

\section{Equations of motion for a winding membrane}

Our starting point is the Nambu-Goto action for a membrane, 
\ba 
{\cal S} =-\mu_2 \int d^3\sigma \sqrt{-\mathrm{Det}(G^{(3)}_{ab})},
\label{memb}
\ea
where $\sigma^a$ ($a=0,1,2$) are the three world-volume coordinates,
$G_{ab}\equiv\partial_a x^A \partial_b x^B g_{A B}$ is the induced
metric on the world-volume, $x^A=(t,\vec{x},y)$ are the space-time
embedding coordinates and $\mu_2$ is the
membrane tension. For a winding membrane in its lowest Kaluza-Klein 
state, we can set $\sigma^2=\theta$ in the Milne metric (\ref{milne11d}).
The action (\ref{memb}) then reduces to 
\be \label{action1}
S=-\mu_2 \int \theta_0|t|d^2\sigma \sqrt{-\mathrm{Det}(G^{(2)}_{\alpha\beta})},
\ee
where now $G^{(2)}_{\alpha\beta}$ is the induced metric on the two-dimensional
world sheet with coordinates $\sigma^0\equiv \tau$ and $\sigma^1\equiv \sigma$. 
Henceforth dots shall denote derivatives with respect to $\tau$ and 
primes derivatives with respect to $\sigma$. 

The action (\ref{action1}) may be viewed as describing 
a string moving in flat space-time with
a tension $\mu_2 \theta_0|t|$ tending to zero as the
string approaches the singularity. Hence one expects all
points on the string to move with the speed of light
in this limit. Equivalently, the same action 
may be viewed as describing a string of fixed tension 
$\mu_2 \theta_0 t_s$, moving in a time-dependent 
background with metric $g_{\mu\nu}=|t/t_s|\eta_{\mu\nu}$.
We shall adopt this latter point of view throughout the paper.

The action (\ref{action1}) is invariant under reparametrizations of
the string world sheet. In particular, we can choose timelike gauge 
$x^0\equiv t=\tau$, and also
$\dot{\vec{x}}\cdot\vec{x}\,'=0$. 
In this gauge, the string
spatial coordinate $\vec{x}(t,\sigma)$ obeys the following
classical equations of motion:
\ba\label{eom1}
\partial_t\big(\epsilon \dot{\vec{x}}\big)=\partial_\sigma
\Big(\frac{t^2\partial_\sigma\vec{x}}{\epsilon}\Big), \qquad
\partial_t\epsilon&=&t\frac{(\vec{x}')^2}{\epsilon}, 
\ea
where $\epsilon$ is an auxiliary quantity (roughly speaking,
the `relativistic
energy density' for the string), defined by 
\be\label{cons1}
\epsilon=\sqrt{\frac{t^2(\vec{x}')^2}{(1-{\dot{\vec{x}}}\,^2)}}.
\ee
Equations  (\ref{eom1}) and (\ref{cons1}) imply that 
\be\label{epseq}
\partial_t(\epsilon^2)=\frac{2}{t}(1-\dot{\vec{x}}\,^2)\, \epsilon^2,
\ee
which will be useful later. Notice in particular that 
as $t$ tends to zero, from (\ref{eom1}) 
$\epsilon$ tends to a constant. Hence from
from (\ref{epseq}) the speed of the string 
tends to unity. 
As pointed out in Ref. ~\cite{turok}, for generic string states, 
equations (\ref{eom1}) are regular for all $t$.

As in flat space-time, it is useful to rewrite these equations in terms
of left and right moving modes, defined by
\be\label{rightleft}
\vec{r}=\dot{\vec{x}} - \frac{|t|}{\epsilon} {\vec{x}}\,', \qquad \qquad 
\vec{l}=\dot{\vec{x}} + \frac{|t|}{\epsilon} {\vec{x}}\,'.
\ee
It is easy to show that $\vec{l}$ and $\vec{r}$ 
are unit vectors, $\vec{r}\,^2=1=\vec{l}\,^2$, as in flat 
space~\cite{kibble}. 
For a closed string, they each describe 
a closed curve on a unit sphere.
Notice also that whereas the timelike gauge we have chosen is
invariant under 
reparameterizations of $\sigma$, the  
the left and right movers are themselves 
reparameterization invariant. Many of the properties of 
oscillating loops can be seen most directly by picturing
the left and right movers as curves on a unit sphere. 
In three spatial dimensions, even in flat space-time
such curves generically cross.
Where this happens the left and right movers coincide, and
it follows that the string moves
at the speed of light for an instant ~\cite{ntnpb}.
As we have already mentioned, in the background of interest, at $t=0$
{\it every} point on the string must move at the speed of light. Hence,
at this moment, the  
the left and right mover curves must actually coincide. 

In terms of the left and right movers, the equations of motion
(\ref{eom1}) read  
\ba
\label{lreom}
\dot{\vec{r}}+\frac{|t|}{\epsilon}\vec{r}\,'+
\frac{1}{2t}\left(\vec{l}-(\vec{r}\cdot\vec{l})
\vec{r}\right)&=&0,\nonumber \\  
\dot{\vec{l}}-\frac{|t|}{\epsilon}\vec{l}\,'+
\frac{1}{2t}\left(\vec{r}-(\vec{r}\cdot\vec{l}).
\vec{l}\right)&=&0.
\ea
In the limit $t\rightarrow 0$, the last terms in each
equation force 
\be\label{coincide}
\vec{r}\rightarrow\vec{l},
\ee
so that, as explained earlier, 
all points on the string reach the speed
of light. 
Conversely, in the limit of large times,
when a loop is well inside
the Hubble horizon, one expects it to evolve as
in flat space-time. As explained below, when we
convert to proper time
and space coordinates, 
the right and left movers 
defined in (\ref{rightleft}) correspond to 
flat space-time 
right and left movers after a reparameterization of $\sigma$.
Our numerical calculations verify that $\vec{r}$ and
$\vec{l}$ tend towards fixed curves at large times, 
providing a straightforward matching onto flat space-time
solutions in this asymptotic regime. 

The equations of motion are non-linear and hard to solve
analytically. We have therefore resorted to a numerical
study of a variety of cases. We have also developed 
analytic approximations in the large time and small
time limits, which we compare to the numerical results.  
Recall that 
the string length is $\sim M_{11}^{-1}$ when we start our
evolution, and the Hubble radius is larger by a factor $\theta_0^{-1}$.
As time runs forward, 
the comoving Hubble radius shrinks as $|t|$ 
and the comoving loop size grows as $|t/t_s|^{-{1\over 2}}$.
Once the string crosses the Hubble radius, its vibrations
cease and the string follows a kinematical super-Hubble 
evolution. 

The `freezing' of the string outside the Hubble radius 
is, as we will discuss below, 
like a sort of measurement process. Depending on the phase
of oscillation of the string, either the string coordinate 
can be frozen, or its momentum. In either case, the conjugate
variable then acquires a large kick following $t=0$. 
After crossing
the singularity, the loop re-enters the Hubble radius and starts 
oscillating as in flat space once again (see Figure \ref{diagram}). 

\subsection{Asymptotic states: flat space description}

At early and late times the string loop is well inside the Hubble
radius and we expect it to follow standard flat space-time evolution,
which we now pause to review. In standard flat coordinates, 
\be\label{flatmetric}
ds_{fl}^2=-dT^2+(d\vec{X})^2,
\ee
we can choose timelike, orthonormal gauge, $T=\tau$ and
$(\partial_T\vec{X})\cdot(\partial_\sigma\vec{X})=0$.
The string then evolves according to the wave equation 
\be\label{flateqns}
\partial_T^2\vec{X}=\partial_\sigma^2\vec{X},
\ee
with the constraint 
\be\label{flatcons}
(\partial_T\vec{X})^2+(\partial_\sigma\vec{X})^2=1.
\ee
The usual left and right movers are given by
\be\label{rightleftflat}
\vec{R}=\partial_T\vec{X} - \partial_\sigma\vec{X}, \qquad \qquad 
\vec{L}=\partial_T\vec{X} + \partial_\sigma\vec{X},
\ee
and the equations of motion may be written
\ba\label{eomsflat}
\partial_T\vec{R}+\partial_\sigma\vec{R}=0, \qquad \qquad
\partial_T\vec{L}-\partial_\sigma\vec{L}=0.
\ea
The solutions may be expressed as a sum over Fourier modes,
\ba\label{fouriermodes}
R^i(T-\sigma)=\sum_{n=-\infty}^{+\infty}
\alpha^i_n e^{-in(T-\sigma)},   
\qquad \qquad
{L}^i(T+\sigma)=\sum_{n=-\infty}^{+\infty}\tilde{\alpha}^i_n
e^{-in(T+\sigma)},
\ea
where reality imposes $(\tilde{\alpha}^i_n)^*=\tilde{\alpha}^i_{-n}$ and 
${(\alpha_n^i)}^*=\alpha^i_{-n}$.
To quantize the string, these Fourier parameters 
are promoted to operators. After following the canonical procedure,
and regularizing and renormalizing the nonlinear constraints, 
one finds there are 
three massless excitations, consisting of states of
the form $\alpha_{-1}^\mu \tilde{\alpha}_{-1}^\nu |0\rangle$ where
$|0\rangle$ is the oscillator vacuum state. These states consist of
a space-time scalar (the dilaton), 
a symmetric and traceless tensor (the graviton) 
and an antisymmetric tensor. These modes are the most relevant
cosmological perturbation theory and evolving perturbations through the
bounce. Hence, the rest of the paper will be dedicated to their
classical analogs. We shall not, however, study the important
issue of mass renormalization, which we defer to future work.

\begin{figure}[t!]
{\centering
\resizebox*{6.27 in}{3 in}{\includegraphics{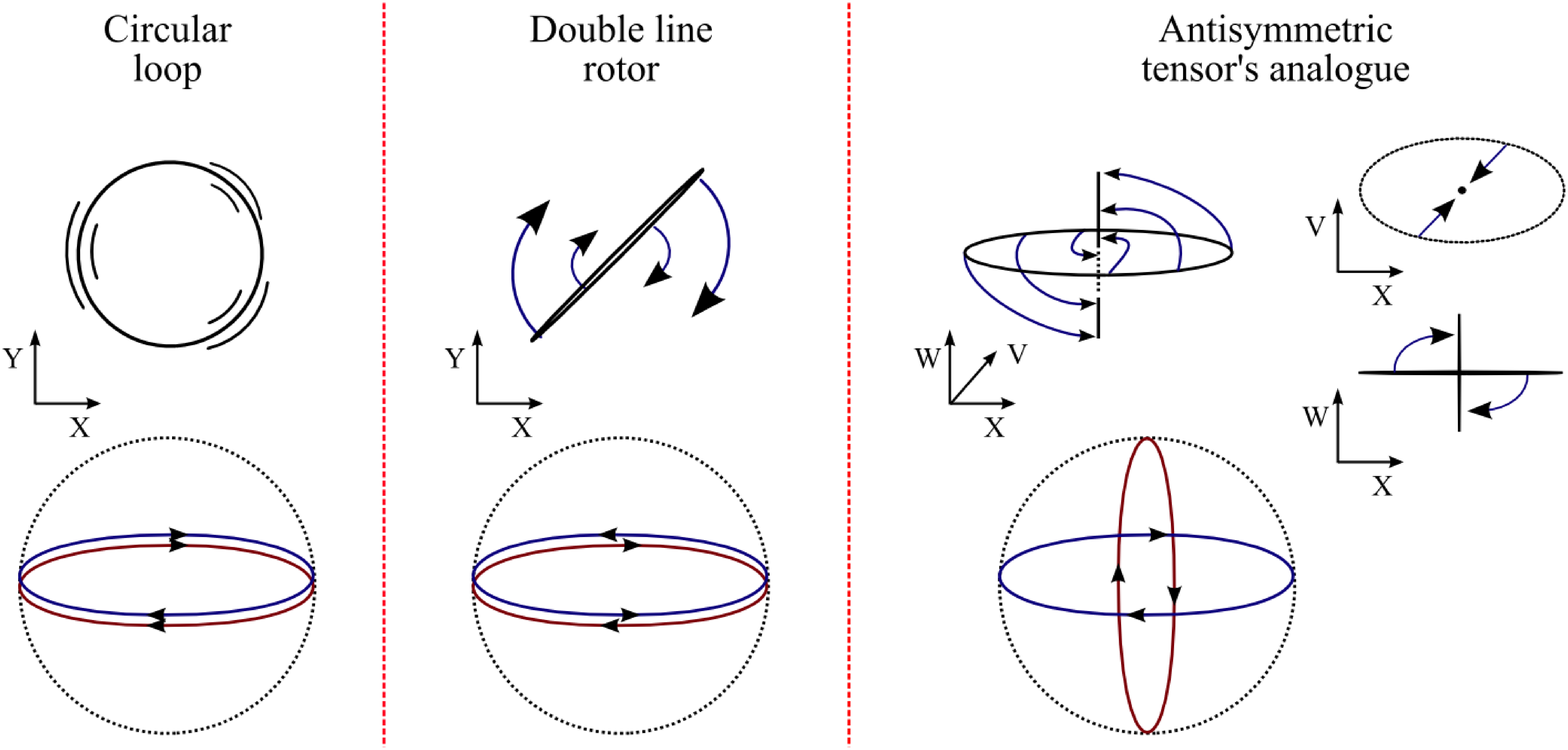}} }
\caption{Classical flat space-time solutions 
consisting of the lowest mode excitations with minimal,
maximal and intermediate angular momentum. 
To get zero angular
momentum the right and left movers lie in the same plane with 
parallel orientation. 
This
configuration corresponds to an oscillating circular loop. The
solution with maximal angular momentum is similar
but has the left and right movers oppositely oriented.
This solution describes 
a spinning doubled line, or rotor. A classical
solution with
intermediate angular momentum is achieved by taking the
right mover circle
to lie in a plane orthogonal to the left mover circle. 
The solution is three-dimensional.
However, under a suitable projection it appears
as an elliptical loop, spinning through an intermediate
doubled-line configuration. 
The $WV$ plane is spanned by $\hat{V}=(0,1,1,0,...,0)/\sqrt{2}$ and
$\hat{W}=(0,-1,1,0,...,0)/\sqrt{2}$.} 
\label{analogues}
\end{figure}

The classical string configuration with no angular momentum, 
hence corresponding to a dilaton-like state,
consists of a circular loop.
In its rest frame, a loop in the $XY$ plane takes the form:
\be
\vec{X}_D=\mathrm{cos}(T)\bigl(\mathrm{cos}(\sigma),
\mathrm{sin}(\sigma),0,\dots,0\bigr), \qquad 0<\sigma \leq 2 \pi.
\label{dil}
\ee
In contrast, the state with maximal angular momentum
for a given energy, analogous to the graviton, takes the
form of a rotor, 
a spinning doubled line:
\be
\vec{X}_G=\mathrm{cos}(\sigma)\bigl(\mathrm{cos}(T),
\mathrm{sin}(T),0,\dots,0\bigr), \qquad 0<\sigma \leq 2 \pi.
\label{grav}
\ee
The left and right movers are easily calculated: they
are circles in the $XY$ plane which are parallel in
the case of the dilaton-like state and antiparallel in the 
graviton-like state. A state of 
intermediate angular momentum may be constructed by taking the
left and right movers to trace out two circles in perpendicular
planes.
Choosing the 
right mover in the $XY$ plane, and the left mover
in the $XZ$ plane, the classical solution is
\ba
\vec{X}_A&=&\frac{1}{2}\Big(\,2\,\mathrm{cos}(T)\mathrm{sin}(\sigma),
\label{flat_anti} \, 
\mathrm{cos}(T-\sigma),\, \mathrm{cos}(T+\sigma),\, 0,\dots ,\, 0\,\Big),
\ea
The three solutions and their description 
in terms of left and right movers on the sphere are depicted 
in Figure \ref{analogues}.

In order to connect these flat space-time solutions to those 
solutions in an expanding universe, we need to relate comoving 
coordinates $\vec{x}$ and conformal time $t$ to physical
coordinates $\vec{X}$ and proper time $T$. Since 
$d\vec{X} = a\, d\vec{x}$ and $dT = a\, dt$, the factors of
the scale factor $a$ cancel out in the velocity. One can also show, 
using the result of Ref. ~\cite{turokv2} that the
time-averaged velocity squared of a loop in its center of
mass frame is ${1\over 2}$,
that (\ref{epseq}) implies that $\int \epsilon d \sigma 
\propto |t|^{1\over 2}
\propto a(t)$. Using this result, one can show using 
an appropriate reparameterization of $\sigma$ that 
the left and right movers (\ref{rightleft})
map precisely onto the flat space left and right movers (\ref{rightleftflat}).
This makes it easy to send in states of any desired
asymptotic form, and to read off the states in which
they come out. A further convenience of the timelike orthonormal
gauge we use is that we do not need to actually perform
the reparameterization of $\sigma$. At any value of $\sigma$ one can
read off the
left and right movers
as functions of the time $t$.

\subsection{Conventions and important quantities for the
  incoming/outcoming modes}

As noted above, we will generally  
choose $\sigma$ to run from 0 to $2\pi$. Moreover, 
in our convention capital letters (like $X$, $E$, etc.) denote 
flat space-time quantities, i.e. those which correspond
to flat space-time variables in the regime where the expansion
of the universe is adiabatic (i.e. where the string loop
is well inside the Hubble radius). 
 On the other hand, lower case symbols 
(like $x$, $\epsilon$, $t$, etc) describe coordinates or 
quantities in the cosmological background.
To specify an incoming mode, for the circle or rotor
we will assume the 
initial string configuration is 
in the $xy$ plane, with some center of mass velocity in the $z$ direction.
We define the latter by
\be\label{n_z}
v_z=\frac{\int \epsilon\, \dot{{z}}\,d\sigma}{\int
  \epsilon\, d\sigma},
\ee
calculated at a time where we start the evolution, 
which we shall formally denote $-t_s$.
The flat space solutions used to specify initial conditions are 
obtained by Lorentz boosting those given above. 
Moreover, we will define the comoving
string size at $t_s$
to be
\be\label{deltadef}
\Delta x=\frac{1}{2\pi}\int d\sigma |\vec{x}\,'|.
\ee
and we shall choose this to be unity in the initial state. 
For each choice of the initial time $t_s$, 
there will be an associated orbifold rapidity $\theta_0 = t_s^{-1}$.
Finally, to get a quantitative measure
of how much energy was produced (or lost) during the $t=0$ transition,
one can define a flat-space-time energy 
\be\label{energy}
E=\int d\sigma\epsilon\, |t_s t|^{-1/2}
\ee 
which approaches a constant for large values of $|t|$, as
explained above. 
The energy ratio between an incoming mode and an outgoing
one ($\eta=E_{in}/E_{out}$) provides a measure of how much energy
was produced during the string's passage across the singularity. 

We will now consider separately the behavior of the 
circle, the rotor and the classical analog of the 
antisymmetric tensor. 

\section{Circular loop}

The simplest nontrivial case to study is a circular loop,
where the symmetry reduces the problem to a single dynamical variable.
Employing the symmetric ansatz 
$\vec{x}(\tau,\sigma)=\rho(\tau)\left(\,\cos(\sigma),\,\sin(\sigma),
\,0,...,0\,\right)$   
where $\rho$ is the comoving radius of the loop,
in the gauge $t=\tau$ the string 
action (\ref{action1}) reduces to
\be
{\cal S}=-2\pi\mu_2\int dt \theta_0|t||\rho|\sqrt{1-\dot{\rho}^2}.
\ee
To minimize clutter, from now on we shall
work in time and
length units in which $2\pi\mu_2 \theta_0 = 1$. 
The canonical momentum $p$ conjugate to $\rho$ and
the Hamiltonian are found respectively to be 
\be\label{eom_circle_b}
p={d \rho\over dt }{|t|S|\rho|\over \sqrt{(1-\dot{\rho}^2)}}, 
\qquad H=\epsilon=\sqrt{p^2+ t^2\rho^2},
\ee
and Hamilton's equations are 
\be
{dp \over dt}=
-{ t^2\,\rho \over H}, \qquad {d \rho \over dt} ={ p\over H}. 
\label{eom_circle_a}
\ee
Finally, because the Hamiltonian is expicitly time-dependent, we obtain 
\be\label{eom_circle_c}
{d H^2 \over dt} =2 t\rho^2.
\ee
\begin{figure}[t!]
{\centering
\resizebox*{3 in}{3 in}{\includegraphics{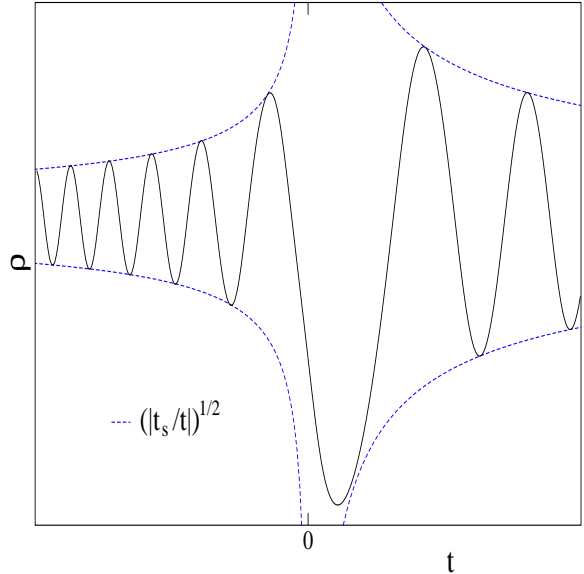}} }
\caption{Numerical solution to equations
  (\ref{eom_circle_b})-(\ref{eom_circle_c}). For large $|t|$, 
the comoving radius $\rho$ 
oscillates, 
with an envelope growing as
$|t_s/t|^{1/2}$. The evolution  
across $t=0$ is smooth and, in this particular example, it results in
a larger amplitude oscillation after the transition. 
} 
\label{dilfig0}
\end{figure}

The qualitative properties of the solutions to 
(\ref{eom_circle_b})-(\ref{eom_circle_c}) are easily seen. 
The energy density $\epsilon=H$ takes its minimal value 
at $t=0$, where the radial speed $d\rho/dt$ reaches unity.
This last effect triggers interesting effects
which we shall describe below. 
The solutions are regular for all $t$ 
except for the special case where the momentum $p$ vanishes at
$t=0$. For that case,
(\ref{eom_circle_c}) implies that 
$H \sim \rho_0 |t|$ at small
$t$. From (\ref{eom_circle_a}) and (\ref{eom_circle_b}), 
one finds that $p$ is very mildly non-analytic, 
$p \sim -{1\over 2} t^2 {\rm sign}(t)$, whereas
$\rho \sim \rho_0 - {1\over 4} t^2 /\rho_0$ is regular. 

Equations (\ref{eom_circle_b})-(\ref{eom_circle_c}) 
are invariant under the rescaling 
$t\rightarrow\Lambda t$, $\epsilon\rightarrow\Lambda^2
\epsilon$, $p\rightarrow\Lambda^2p$ and $\rho\rightarrow\Lambda \rho$,
hence solutions for loops of different sizes are trivially
related, and inequivalent solutions are labelled by only one 
parameter, which may, for example, be taken to be
the asymptotic phase of the oscillation.

The motion of the loop is simplest to describe in two
asymptotic regimes, corresponding to large and small 
$|t|$, when the 
loop's radius $\rho$ is well within, or outside, the 
Hubble radius.
In the first regime the loop oscillates with fixed 
amplitude and period in proper time. In 
comoving coordinates
and conformal time the oscillation amplitude 
changes as $a(t)^{-1}=(t_s/|t|)^{1/2}$, and the frequency
changes as $a(t) \propto (|t|/t_s)^{1/2}$,
as illustrated in Figure
\ref{dilfig0}. 
The energy density 
$\epsilon$ scales as $a(t)=(|t|/t_s)^{1/2}$ in this regime.

\begin{figure}[t!]
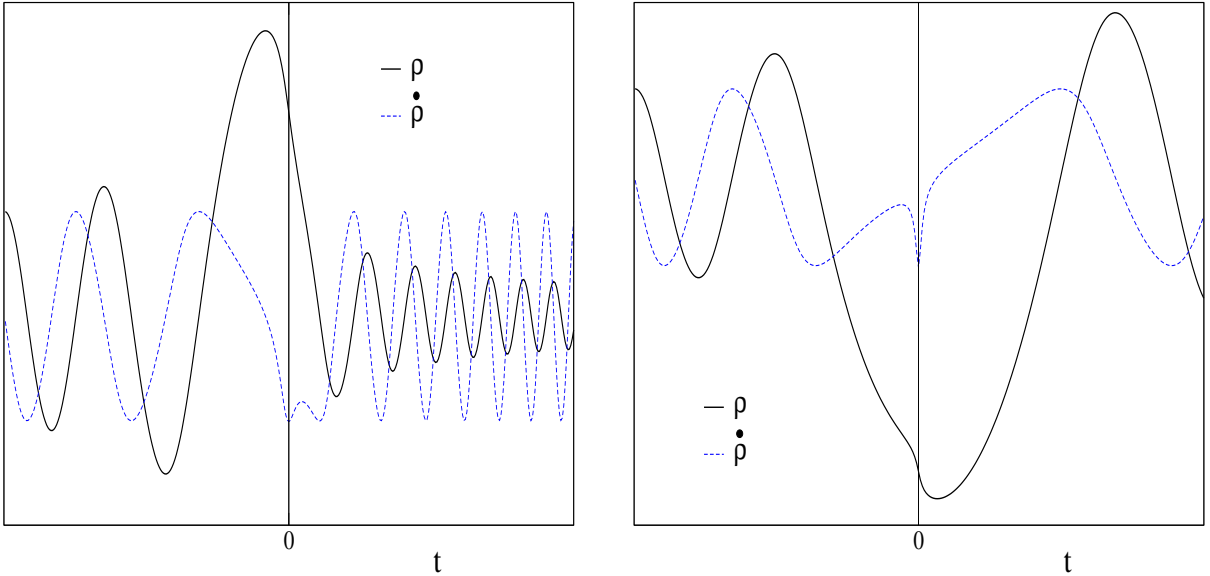

{\centering
\resizebox*{3in}{3in}{\includegraphics{dilaton1a_no.eps}} 
\hspace{.5cm}
\resizebox*{3in}{3in}{\includegraphics{dilaton2a_no.eps}} }
\caption{The radius (solid black line) and the velocity (blue dashed
  line) are plotted for two different circular loops. In the left
  plot the singularity is crossed after a maximum in the radius and
  almost at a minimum in the velocity. In this particular case, there
  is a loss of energy after the transition: the outgoing mode
  has a smaller amplitude than the incoming one. The right plot shows
  the opposite situation, with the radial coordinate approaching 
  a maximum
  at the singularity, and energy being produced.
}
\label{dilfig2}
\end{figure}

Once the comoving Hubble radius $|t|$ falls below the loop radius $\rho$,
the loop enters a new kinematical regime in which the
tension plays a subdominant role.
As the time tends to zero, all points on the string 
approach the speed of light. The loop receives a `kick' from
passing through $t=0$ and emerges with a shift in its 
its oscillation phase and a net energy gain or loss.
In the case of a circular loop, if the radius
is expanding at $t=0$ the loop 
gains energy, whereas if it is contracting the loop loses energy.
Time reversal, which is a symmetry of the equations,
relates these two situations. Figure
\ref{dilfig2} shows two examples.

\subsection{Asymptotic states at large $|t|$}
\label{asymst}

\begin{figure}[t!]
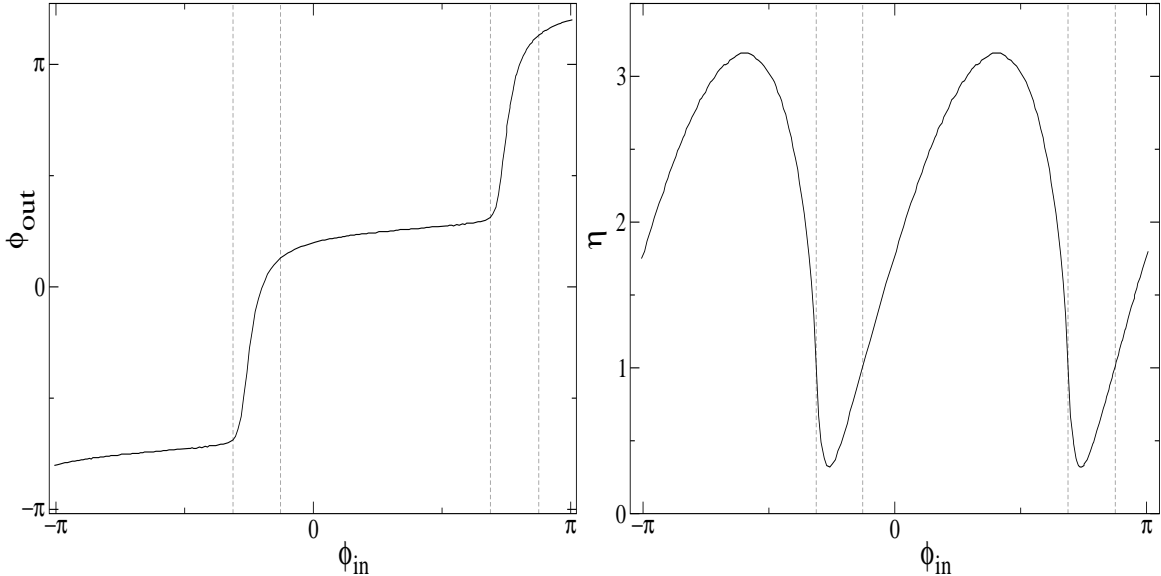

{\centering
\resizebox*{3in}{3in}{\includegraphics{phin_phout.eps}}
\resizebox*{3in}{3in}{\includegraphics{phin_eta.eps}} }
\caption{The outgoing phase $\phi_{out}$ measured at large times
  is shown as a function of the incoming phase $\phi_{in}$ (left plot).
  There are two different regions, one where $\phi_{out}$ varies steeply and
  the other where it is almost flat. The right hand plot shows the
  energy production ratio $\eta=E_{out}/E_{in}$ as a function of
  the incoming phase $\phi_{in}$. The ratio $\eta$ is less than one
in the steep region of the left hand plot, and greater than one outside it. Two
  neighboring regions are related by time-reversal 
  (which sends $\eta$ to $1/\eta$),
  and they are separated by special points at which $\eta=1$.
 These special points, denoted by the
  vertical dashed lines, correspond to the loop
  crossing $t=0$ with either zero momentum or zero radius.}
\label{phases}
\end{figure}

In order to parameterize the incoming and outgoing states, 
it is helpful to perform a canonical transformation to new
coordinates
($\phi,\,\mathcal{P}$):
\be
\label{newdef}
\mathcal{P}=\frac{p^2+ t^2\rho^2}{2|t|}, 
\qquad\qquad \tan(\phi)=\frac{\rho |t|}{p}.
\ee
At large times, the angle $\phi$ represents the oscillation 
phase while
the conjugate momentum $\mathcal{P}$ tends to a constant 
measuring the energy stored in the loop.
The Hamiltonian equations now read
\ba\label{eom_can}
{d {\mathcal{P}}\over dt} &=&-\frac{\mathcal{P}}{t}\cos(2\phi), \nonumber\\ 
{d {\phi}\over dt} &=&\sqrt{\frac{|t|}{2\mathcal{P}}}+\frac{1}{2t}\sin(2\phi).
\ea
These equations are readily solved at large times, giving 
\be
\mathcal{P}\rightarrow \mathcal{P}_\infty=\mathrm{const.}, \qquad
\qquad \phi\rightarrow \int \sqrt{\frac{t}{2\mathcal{P}}}dt\sim
\pm \frac{2}{3}\sqrt{\frac{|t|^3}{2\mathcal{P}_\infty}}+\phi_0, 
\ee
where $\phi_0$ is a constant phase. From (\ref{newdef}) it
follows that the energy (\ref{energy}) 
$ E\propto \sqrt{\mathcal{P}}$ and hence it 
tends to a constant at large times, as expected. 
In the usual comoving $\rho$ coordinate, the asymptotic behavior 
of the solution at large times is
\be
\rho\sim \sqrt{\frac{2\mathcal{P}_\infty}{t}}
\sin\left(\frac{2}{3}\sqrt{\frac{t^3}{2\mathcal{P}_\infty}}+\phi_0\right).
\ee
Upon identifying proper time and radius, this takes the form of the 
flat space solution (\ref{dil}), up to an arbitrary phase. 
Due to the rescaling symmetry discussed above, the
only nontrivial parameter in the incoming state is the phase
$\phi_0=\phi_{in}$. The outgoing state may be completely characterized by
a similar phase $\phi_{out}$, and by the energy ratio
$\eta=E_{in}/E_{out}$, both of which can be expressed as a function 
of the incoming phase. Figure \ref{phases} shows all the information
needed to describe the classical transition, {\it i.e.}, the
outgoing energy and phase in term of similar ingoing quantities. 
In the case of 
the energy ratio $\eta$, one can determine the 
energy production for an incoming mode with fixed amplitude 
but unknown
phase, or equivalently,
the average energy production weighted by the Liouville measure.
By integrating the curve in Figure \ref{phases}, we find
\be
<\eta>=\frac{1}{2\pi}\int_0^{2\pi}\eta \ d\phi_{in}= 2.12.
\ee

Before moving into the regime close to the singularity let us briefly 
comment on the consequences of including a non-zero center 
of mass velocity. As we increase the initial center of mass velocity $v_z$,
the energy production decreases so that the plot for $\eta$ (Figure
\ref{phases}) shrinks in the vertical direction. 
This is to be expected since the
$z$ component of the string's velocity means that it is closer
to the speed of light and hence suffers less
of a `kick' as it passes through $t=0$. This translates
into less energy production or energy loss. 

\subsection{Behavior near the singularity}

Classically, the state of a circular loop with zero
center of mass velocity is specified by two numbers,
its radius and the canonically conjugate momentum. However because
of the scaling symmetry discussed above, there is
really only one physical parameter. Near the singularity,
we can choose the scale-invariant combination $|p_0/\rho_0^2|$, where $p_0$
and $\rho_0$ are the values of $p(t)$ and $\rho(t)$ at $t=0$. This
combination compares the radial 
momentum to the size of the loop at the singularity:  if
$|p_0/\rho_0^2|\ll 1$, then the radius is large and not much
changed during the transition across the singularity. 
In contrast, if $|p_0/\rho_0^2|\gg 1$,
then the momentum is hardly changed during the transition. 
In each case, the conjugate variable undergoes a large change
across the transition. 
Heuristically, one can think of the
transition as `measuring' the loop radius or its momentum in the
two cases, so that the conjugate variable acquires a large jump,
as a consequence of Liouville's theorem. 
Figures 6 and 7 illustrate
the two situations. The former situation, where the radius is
`frozen' during the transition, describes the rising portion
of the $\eta (\phi_{in})$ curve in Figure 5. As can be seen
from the plot, this is the more common situation for 
initial states with uniformly chosen $\phi_{in}$.

Likewise, certain features of Figure 5 are readily understood.
Recalling that the
phase  $\phi$ and the `energy'  $\mathcal{P}$
defined in (\ref{newdef}) are canonically conjugate, one sees
how the squeezing of the outgoing phase causes the loop energy
to be amplified, and vice versa. 
\begin{figure}[t!]
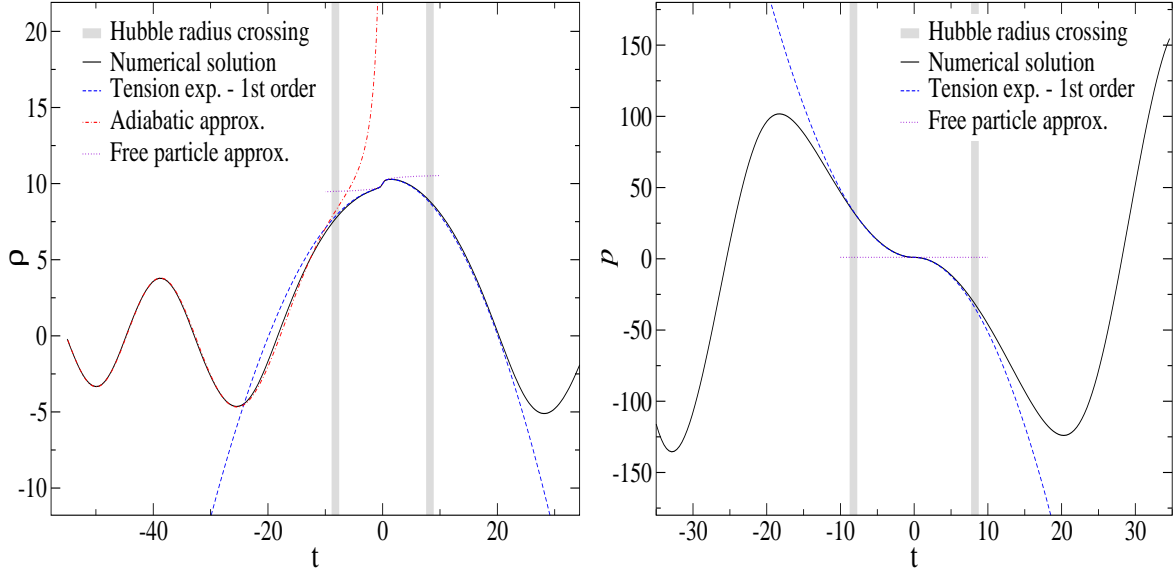

{\centering
\resizebox*{3in}{3in}{\includegraphics{dil-s-exp3a.eps}} }
{\centering
\resizebox*{3in}{3in}{\includegraphics{dil-s-exp3b.eps}} }
\caption{Both numerical simulations and analytical approximations
for the radius $\rho$ (left plot) and the canonical momentum (right
plot) are shown as a function of time for a circular loop. 
Due to the small invariant parameter $|p_0/\rho_0^2|=0.1$, it
is the loop radius (left plot), not the momentum (right plot)
that `freezes' during the transition.
The expansion in the string tension (see text)
works well in this case,
even beyond Hubble crossing.
In contrast, the free particle 
approximation only agrees with the numerics
close to $t=0$, where there is drastic change in the
string momentum. If the loop radius is taken to be
of order the string length $l_s$ at a time $t_s$, then 
this case corresponds to $\theta_0\sim 1/260$.}
\label{dil-s-exp}
\end{figure}
\begin{figure}[t!]
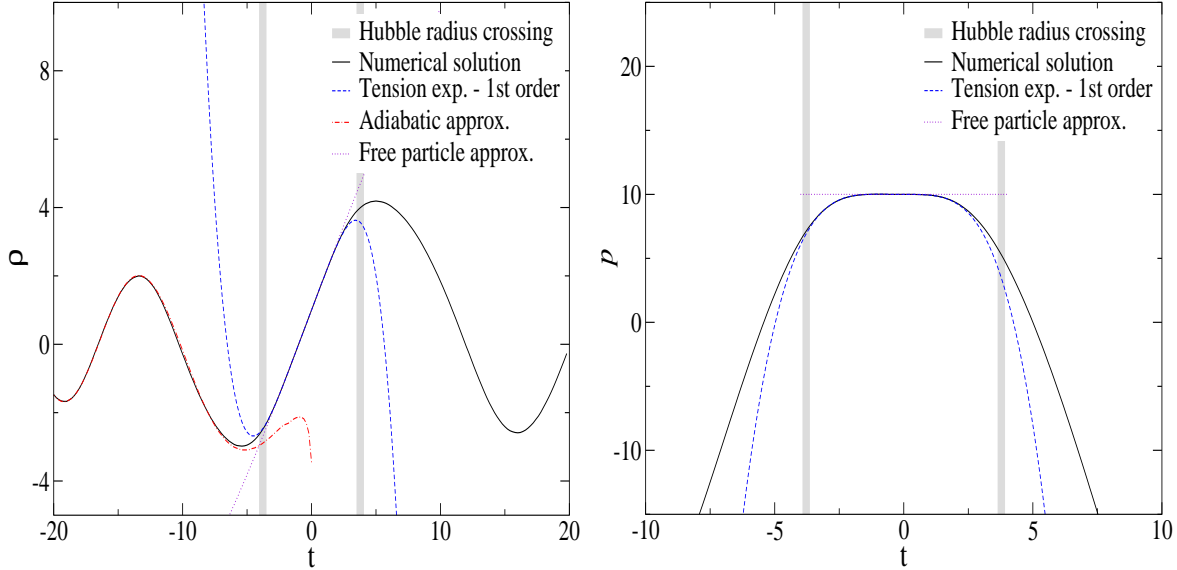

{\centering
\resizebox*{3in}{3in}{\includegraphics{dil-s-exp2a.eps}} }
{\centering
\resizebox*{3in}{3in}{\includegraphics{dil-s-exp2b.eps}} }
\caption{Another circular loop example, where in contrast
to the situation in Figure 6, 
the ratio $|p_0/\rho_0^2|=10$ so that the momentum $p$
is nearly frozen across the transition, whereas
the radius $\rho$ changes significantly - undergoing 
a complete oscillation. In this case, the expansion in the string tension
works only until Hubble crossing, and 
 describes the maximum and minimum with less precision. 
If the loop radius is taken to be
of order the string length $l_s$ at a time $t_s$, then 
this case corresponds to $\theta_0\sim 1/128$.
} 
\label{dil-s-expb}
\end{figure}

\subsection{Analytical treatment: an expansion in the string tension}
\label{lamsec}

To understand the behavior near the singularity better, 
we consider expanding about $t=0$. 
The simplest possibility is merely to perform 
a Taylor series in $t$, 
but this expansion is poorly convergent. Instead, we have
developed an analytic solution method which is formally
an expansion in the string tension, or $1/\alpha'$. 

To perform this expansion, let us introduce a formal parameter 
$\lambda$ multiplying the tension term
in the Hamiltonian, so that
\be\label{ham}
H \rightarrow \sqrt{p^2+ \lambda t^2\rho^2}. 
\ee
We now calculate the Hamiltonian equations as before, using
(\ref{ham}) to eliminate the $\lambda$-dependence of the 
equation for the time-dependence of the Hamiltonian. We
obtain
\be
{dp \over dt} =
-\lambda { t^2\,\rho\over H},  \qquad {d \rho \over dt} ={ p\over H},
\qquad {d H^2 \over dt} ={2 \over t} (H^2-p^2).
\label{S-circle}
\ee
In this form, $\lambda$ appears in only one term 
in the equations of motion. We can solve the equations
as a power series in $\lambda$ by 
inserting 
the lower order solution into the term involving $\lambda$ at each new
order.
At the end of the calculation, we set $\lambda=1$
and fix all remaining integration constants. Notice that 
we do not allow
powers of $\lambda$ in the initial conditions: the
approximation applies only to the dynamical evolution.

The zeroth 
order solution 
is found by setting $\lambda=0$ in equations (\ref{S-circle}).
The first equation implies 
$p=p_0$, a constant. The last equation 
then implies $H=\sqrt{p_0^2+C^2 t^2}$, with $C$ a constant.
Finally, the equation for $\rho$ is easily integrated. 
Finally,
we set $\lambda=1$ and identify the integration constant $C$.
Comparing the expression for
$H$ in our solution 
with the exact Hamiltonian, we can see that 
$C=\rho_0$, the radius of the loop at $t=0$.
Hence we obtain the zeroth order solution
\be\label{free_part}
\rho=\rho_0+\frac{p_0}{\rho_0}\sinh^{-1}
\Big(\frac{\rho_0}{|p_0|}t\Big).
\ee
This solution is the same as that for a
winding string, behaving as a particle of mass
$M \propto |t|$, which reaches the 
speed of light instantaneously at $t=0$ as
described in Ref.~\cite{turok}.

It is straightforward to compute the solution 
to higher order in $\lambda$, although the 
integrals become increasingly difficult.
Details are given in 
Appendix \ref{appex}. 
In general, the series solution for $\rho$ and $p$ to a given order
are truncated polynomials in $p_0/\rho_0^2$, where the
coefficients are functions of $\chi\equiv\sinh^{-1}
\Big(\frac{\rho_0}{|p_0|}t\Big)$. The series converges for much
larger times than a simpler Taylor series in $t$ in part because
$\chi$ only grows logarithmically with time.
If $p$ consists of a 
polynomial of order $n$ in $p_0/\rho_0^2$, then $\rho$ consists
of terms up to order $n+1$.
Therefore, if $|p_0/\rho_0^2|\gg 1$ the higher power terms will be more important
than the lower power ones, and vice versa for
$|p_0/\rho_0^2|\ll 1$. To second or higher orders in the expansion, the
time-dependent series coefficients include 
polylogarithm functions, which appear 
naturally in Feynman diagrams
~\cite{kreimer}, and in number theory. This may be a sign of deeper
underlying simplicity.

In Figures 6 and 7, the expansion in the string tension, taken to
second order, is compared with
numerical solutions. The analytic approximation shows good agreement
with the numerics up to times where the loop is starting to be 
well described by the appropriate adiabatic flat-space oscillatory solutions,
{\it i.e.}, when the loop is inside the Hubble radius.
We conclude that the expansion in 
the string tension which we have defined
is a powerful tool for studying classical evolution 
right across the transition. 
\begin{figure}[t!]
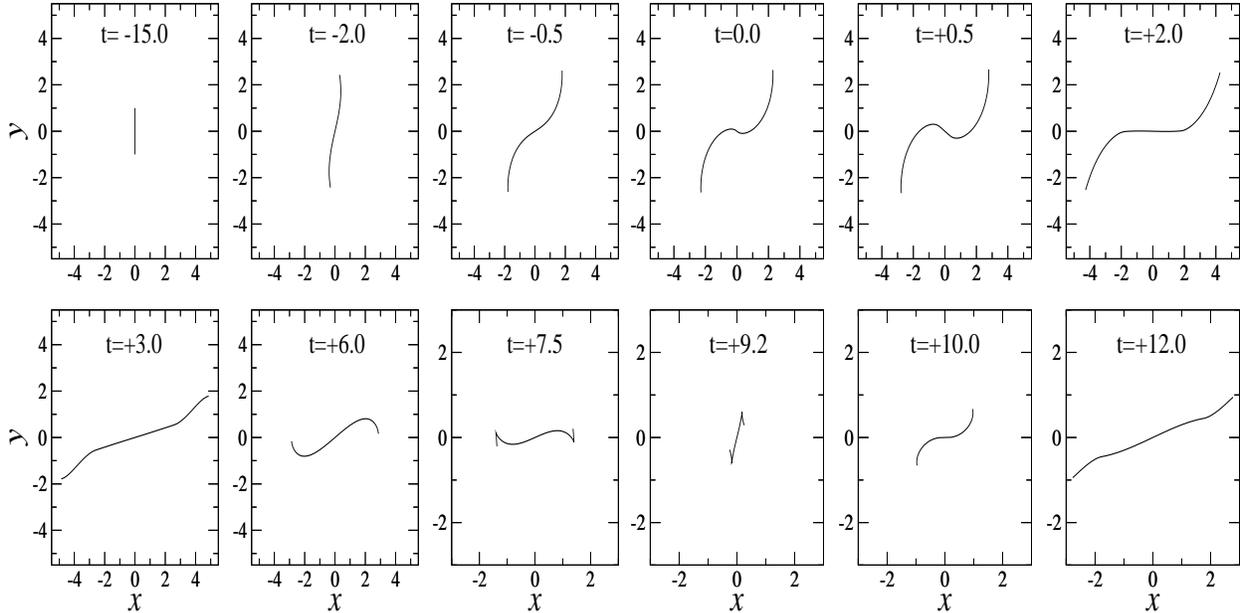

{\centering
\resizebox*{3.2in}{3.2in}{\includegraphics{rotor-pos1.eps}}
\resizebox*{3.2in}{3.2in}{\includegraphics{rotor-pos2.eps}} }
\caption{An incoming flat space-time rotor solution (\ref{grav}),
with a small 
initial center of mass velocity, $v_z=0.01$ and $t_s=25$. 
The rotor evolves into an ``S'' shape in the $XY$ plane 
as the two opposite arms speed up towards the speed of light.
The outgoing state involves a superposition of higher oscillation
modes of the string, as may be seen from its evolution.
}
\label{sshape}
\end{figure}

\section{Rotor}

The classical analogue of the graviton is a rotor
solution, a spinning doubled line whose ends
move at the speed of light. The flat space-time
solution representing this incoming state is 
given by equation (\ref{grav}) or, if it
has nonzero center of mass velocity,  a Lorentz-boosted version
of it. The appropriate initial conditions for the 
time-dependent background we study are given by 
computing the flat space-time left and right movers, and translating these
into expanding universe left and right movers at the initial time. 
The solutions take the form of a doubled
line for all time and the 
two end-points always move at
the speed of light. The evolution is more complicated
than that for a circular loop because there is a non-trivial
dependence on $\sigma$. 

There is a set of measure zero on the space of initial 
conditions where the evolution is singular, when the
center of mass momentum of the rotor is precisely
zero. Since there is no 
preferred direction orthogonal to the plane of the rotor,
its central point must remain at rest 
for all $t$. This conflicts with the requirement
that all points must move at the speed of light at $t=0$, 
hence the
solution must go singular. However, we do not believe
this will cause any problem in the quantum theory. States
of zero momentum are of zero measure on phase space. 
So we shall study the behavior of a rotor with 
nonzero center of mass momentum $p_z$, in
the limit
as $p_z$ tends to zero. We shall find that the
resulting non-analyticity in the classical
solution is rather mild, and thus likely to 
overwhelmed by the measure in any physically
realistic calculation. 
\begin{figure}[t!]
{\centering
\resizebox*{3.2in}{3.2in}{\includegraphics{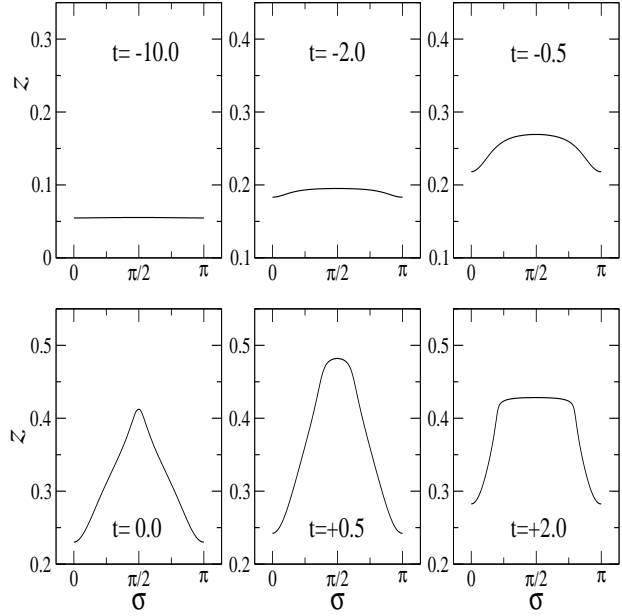}} }
\caption{The profile of the incoming rotor shown in
Figure \ref{sshape}, in the $z$ direction. The passage 
through $t=0$ results in a kink being produced 
at $t=0$, which then propagates outwards across
the string.
} 
\label{kink}
\end{figure}

By boosting the static solution
(\ref{grav}) we can obtain a solution with 
arbitrary $v_z$, as defined in (\ref{n_z}). 
Any simulation is then characterized by the
starting time $t_s$ and the initial center of mass
velocity $v_z$. 
Contrary to the circular loop,
the oscillation phase of the incoming rotor 
is of no physical significance since it can be removed by
a spatial rotation. Therefore, as far as
the classical dynamics is concerned, the solution
for a rotor depends only upon $v_z$ and there
are no other parameters to consider. 

The evolution of such a moving
rotor across $t=0$ is illustrated in Figures 
\ref{sshape} and
\ref{kink}.
In the $XY$ plane, it develops an `S' shape. This 
may be understood as a consequence of 
the opposite arms of the rotor speeding up to
the speed of light in opposite directions
as $t=0$ approaches. In the $Z$-direction, as
the central
point speeds up to the speed of light it creates a 
kink which then runs out across the string. 
\begin{figure}[t!]
{\centering
\resizebox*{3.5in}{3.5in}{\includegraphics{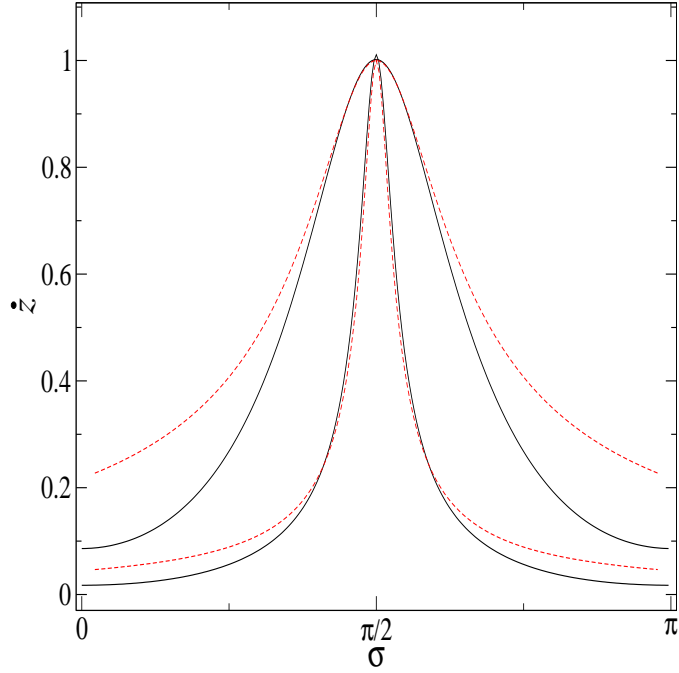}} }
\caption{The kink in the transverse velocity $\dot{z}$ around the middle
  of the rotor, for two very small velocities. The profile (\ref{spike})
 fits the shape very well for small transverse velocities (red
  dashed line), where the quantity $<\dot{z}_0>\equiv\int \epsilon_0 \dot{z}_0 
  d\sigma/\int \epsilon_0 d\sigma$ measures the center of mass
 velocity at $t=0$ in the $z$ direction. The initial conditions were:
 $t_s=5$, $v_z= 0.05$ for the wider curve and $v_z=0.01$ for the narrower
curve.}
\label{kinkzdot}
\end{figure}

\subsection{Profile of the Kink}

As the rotor approaches $t=0$, it develops a kink in the
direction of its motion, whose size and shape 
depends on the magnitude of the transverse momentum. For 
zero transverse momentum, the classical evolution is
ill-defined. This is a set of measure zero, but the 
breakdown of the classical equations at $p_z=0$ may be
indicative of some divergence. Therefore it is important 
to study the evolution for small $p_z$ to see whether
physical quantities diverge that limit. 

As the transverse momentum gets smaller, 
the kink gets narrower and narrower, as shown in 
Figure \ref{kinkzdot}. 

\begin{figure}[t!]
{\centering
\resizebox*{3.2in}{3.2in}{\includegraphics{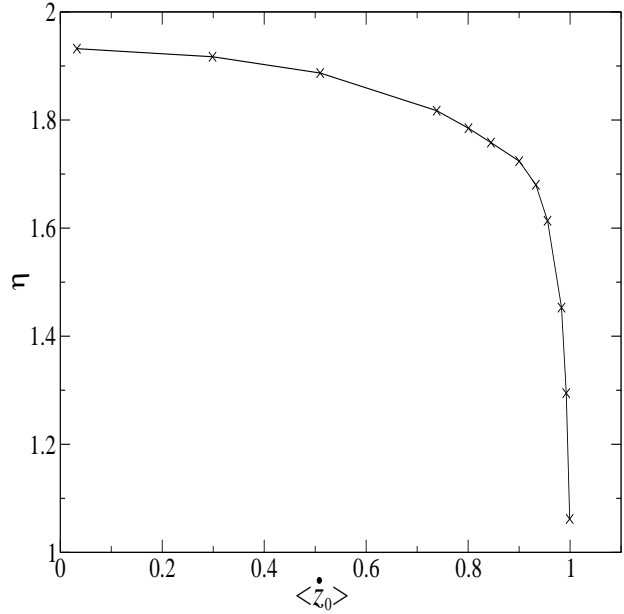}} }
\caption{The energy ratio
  $\eta=E_{out}/E_{in}$ for a rotor is plotted against the average 
center of mass  speed
of the loop 
  at $t=0$, {\it i.e.}, $<\dot{z}_0>\equiv\int \epsilon_0
  \dot{z}_0 d\sigma/\int \epsilon_0 d\sigma$, showing that
the net energy production is unaffected by the non-analytic
spike produced in the limit of small transverse momentum.
} 
\label{partprod_grav0}
\end{figure}

The profile of the kink may be modeled analytically as
follows. 
Consider the equation
for the velocity of the string in the $z$ direction, $\dot{z}$.
From equation (\ref{eom1}), at $t=0$ we have
\be
\dot{z}=\frac{\pi_z(0)}{\epsilon(0)}=\frac{\pi_z(0)}{\sqrt{\vec{\pi}\,(0)^2}},
\ee
where $\vec{\pi}= \epsilon \dot{\vec{x}}$ is the canonical
momentum density at each point on the string, and we used 
(\ref{cons1}) to get the last equality. For
small transverse velocities, near the center of the rotor we can replace
$\pi_z\sim<\dot{z}_0>\equiv\int\epsilon_0\dot{z}_0d\sigma/\int
\epsilon_0 d\sigma$ and use the Taylor expansion
$\pi_x^2+\pi_y^2\sim(C\,\sigma)^2$, with $C$ a constant. Therefore,
near the center of the rotor, the kink's profile at $t=0$ is 
approximated by
\be
\label{spike}
\dot{z}=\frac{1}{\sqrt{1+\left(\frac{C\,\sigma}{<\dot{z}_0>}\right)^2}}.
\ee
This fit works well against our numerical results, 
as shown in Figure
\ref{kinkzdot},
with $C^2\sim 0.2$.
As we take the centre of mass momentum of the loop to zero, 
the kink becomes more and more strongly 
localized and involves higher and higher oscillation modes
of the string. However, these do not contribute significantly to
the energy. 
Figure \ref{partprod_grav0} shows $\eta = E_{out}/E_{in}$ as a 
function of the loop momentum. In the limit of
small momentum, $\eta$ tends to a finite constant
$\sim 1.93$, showing that there is little energy associated with 
the spike generated at the center of the rotor. 
We conclude that at low momenta, we produce a 
a non-differentiable but
finite spike, although
physical quantities like the 
energy or momentum remain perfectly finite.

\subsection{Expansion in the String Tension}

Just as we have done for the circular loop, we can 
describe more general string states, including 
the rotor, using a formal expansion in 
the string tension. The details are given in Appendix \ref{appex}.
Following the same general method as given 
for the circular loop, the zeroth order solution is given as
\be\label{freepartsol}
\vec{x}=\vec{x}_0+\frac{\vec{p}_0}{|\vec{x}'_0|}
\,\mathrm{sinh}^{-1}\bigg(\frac{|\vec{x}'_0|}{|\vec{p}_0|}t\bigg),
\ee
where $\vec{p}_0$, $\vec{x}_0$ and $\vec{x}'_0$ are functions of $\sigma$.

Higher terms in the 
expansion are more difficult to calculate and
the integrals can only be done numerically 
(see Appendix \ref{appex} for details). Figure \ref{grav-s-exp} shows the
numerical solution plotted 
against the zeroth order term in the expansion, the first order term,
and the flat space approximation, respectively. The comparison is
made for an end point of the rotor (right plot) and a
representative point further in (left plot). For the former, 
the adiabatic approximation holds very accurately 
up to Hubble crossing. Since the point is moving at the speed
of light, the zeroth and first order terms 
of the expansion are nearly
identical, providing a good approximation to the motion
outside the Hubble radius. At Hubble re-entry, the
solution reverts to a form close to flat-space evolution
but this is more complex to determine hence we have not
attempted to graph it. For generic points on the rotor, 
the adiabatic approximation is less accurate at earlier times:
the speeding up of the central point as the kink is created
in effect extracts energy from the normal spinning motion.
As can be seen from the diagram, the first order expansion
in the string tension does reasonably well in modeling the
behavior around $t=0$ up to the first turning point.

\begin{figure}[t!]
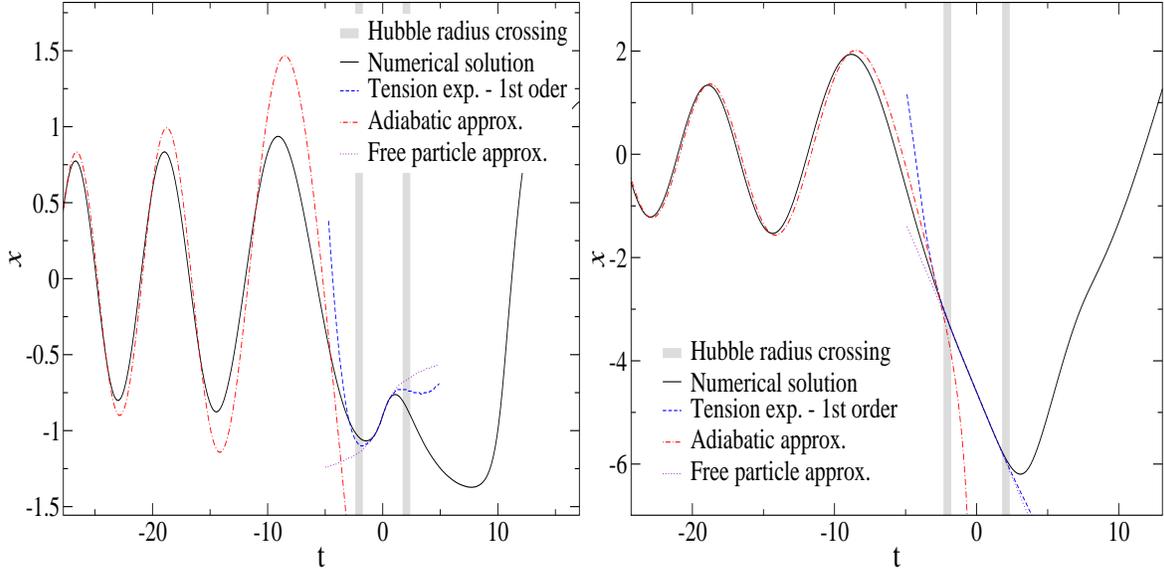

{\centering
\resizebox*{3.in}{3.in}{\includegraphics{grav-s-exp3.eps}}
\resizebox*{3.in}{3.in}{\includegraphics{grav-s-exp3b.eps}} }
\caption{The numerical solution for two points on a rotor. The left
plot shows the $x$ coordinate at 
$\sigma=\pi/4$ and the right plot shows the $x$ coordinate
of the end point $\sigma=0$. The adiabatic
flat space approximation reproduces the numerical 
solution reasonably well until the loop crosses the 
Hubble radius. Beyond this point, an 
expansion in the string tension becomes
a much better description. In this plot, the initial 
center of mass speed was $v_z=0.05$ and $t_s$ was taken
as $35$.
 }
\label{grav-s-exp}
\end{figure}
\begin{figure}[t!]
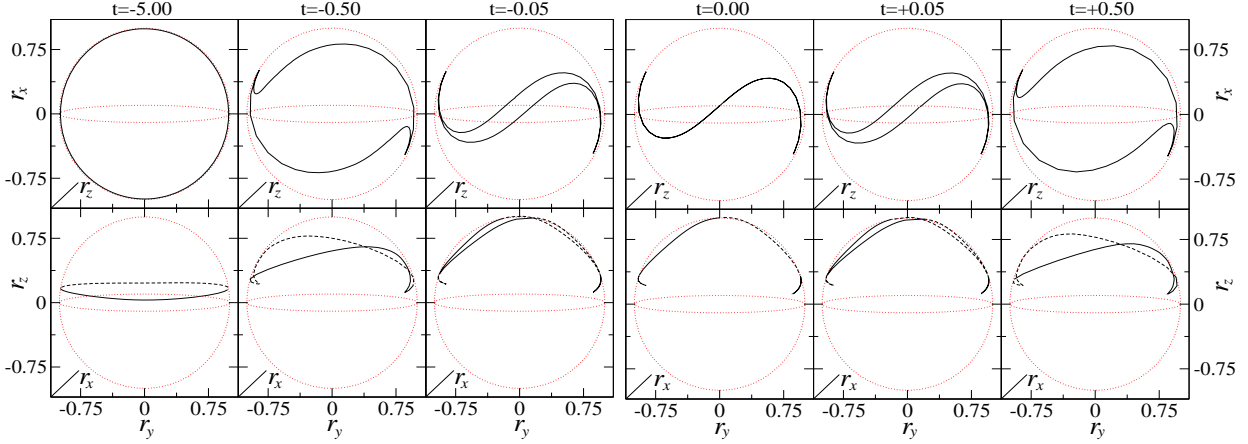

{\centering
\resizebox*{3.2in}{2.3in}{\includegraphics{sphere1.eps}}
\resizebox*{3.2in}{2.3in}{\includegraphics{sphere2.eps}} }
\caption{The evolution of the left mover $\vec{l}$ for 
an incoming rotor. The left 
mover defines a closed curve on a three-sphere. The right mover
follows the same curve but with the opposite orientation.
The condition (\ref{coincide}) that the left and right 
movers coincide $t=0$, 
means that the evolution must deform the trajectories over
the north pole of the three sphere. 
The only way to get the two curves to
coincide is by deforming them into a double line
at $t=0$, running directly
through the north pole (or the south pole if $v_z$ is
negative). 
The initial parameters
here were $v_z=0.05$, $t_s=45$.
}
\label{spheremovers}
\end{figure}
\begin{figure}[t!]
{\centering
\resizebox*{3.2in}{3.2in}{\includegraphics{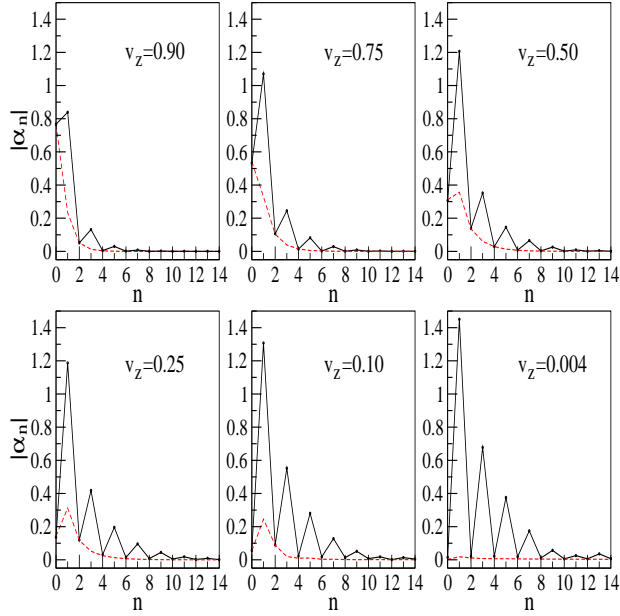}}}
\caption{Fourier amplitudes of outgoing right movers calculated in the
asymptotic adiabatic regime for different initial transverse speeds.
The red dashed line shows the $z$ component 
only, hence $\alpha^z_0$ (the $n=0$ value of the red dashed
line) is the final speed $v_z$.
By symmetry, $\alpha^x$ and $\alpha^y$ have only 
odd-$n$ modes whereas  
$\alpha^z$ has both odd and even $n$ components. In these plots
$t_s=25$.
}
\label{fourier2}
\end{figure}
\begin{figure}[t!]
{\centering
\resizebox*{3.2in}{3.2in}{\includegraphics{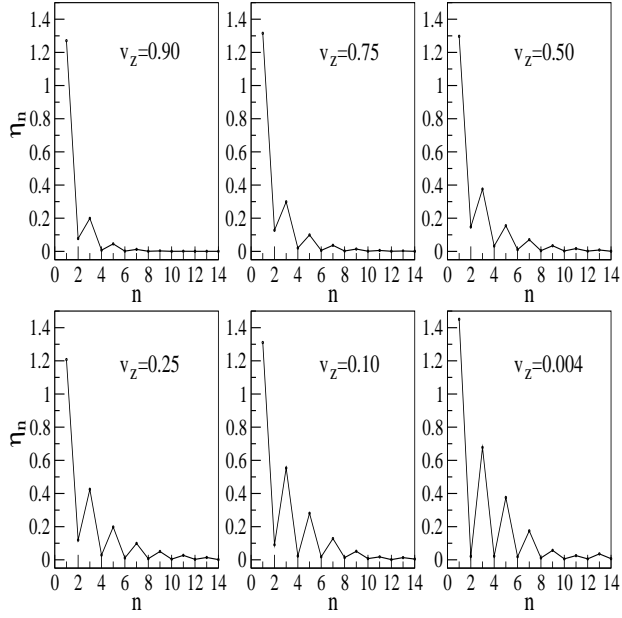}} }
\caption{The energy ratio $\eta$ for each mode number $n$, representing
the amplitude of each outgoing mode divided by the amplitude
of the incoming mode. Larger initial speeds $v_z$ generate
smaller amplitudes in the higher modes. 
The decay of $\eta$ with $n$ resembles the plane wave
estimate ($\sim e^{-\beta\sqrt{n}}/\sqrt{n}$) found in ~\cite{tolley}
for high velocities, but at low velocities $\eta$ decays more rapidly
than expected from that work, and in particular 
the production of high $n$
modes remains finite.}
\label{fourier3}
\end{figure}

\subsection{Outgoing state}

When the rotor is evolved 
forward to large times, the outgoing state consists
of a complicated mixture of 
higher oscillation modes. The precise mixture is 
most easily understood by
representing the outgoing solution in 
left and right moving
modes, which tend to fixed curves on the unit sphere
at late times. 
One can also use the right and left movers 
to understand the motion of the rotor 
across the singularity 
(Figure \ref{spheremovers}).
The two curves start out near the equator
(for small $v_z$) and oppositely oriented. 
For zero center of mass velocity, the 
the curves are confined to the $xy$ plane and there is no
way for them to coincide at $t=0$. However, if $v_z$ is 
positive (or negative), the curves sweep over
the north (or south) pole and coincide at $t=0$.
 
To quantify the level of excitation produced in the
passage across 
$t=0$,  we can track the 
the evolution of the right and left movers in time
at one particular value of $\sigma$ on the string. 
For very large times, the right and left
movers look like flat space solutions and become periodic
functions of proper time. Once the loop is
well inside the Hubble radius, for a single period the
difference between conformal time and proper time is
negligible. Hence we can just choose one value of $\sigma$
and follow the evolution of left and right movers there.
We write the 
expansion of the left and right movers (\ref{fouriermodes}) as
\ba\label{fouriermodes2}
r^i=\sum_{n=-\infty}^{+\infty} \alpha^i_n e^{-in (t-\sigma)},
\qquad \qquad
l^i=\sum_{n=-\infty}^{+\infty}\tilde{\alpha}^i_n e^{-in (t+\sigma)},
\ea
and compute the 
$\alpha^i_n $ and $\tilde{\alpha}^i_n$ by Fourier transforming
the solution with respect to $t$ over a single period. As is seen 
in Figure 19, identifying the periodicity is straightforward. 
As the initial center of mass velocity $v_z$ is tuned down,
modes at higher and higher $n$ are excited, although as 
shown above,
the total energy in the outgoing string converges to a 
finite limit as $v_z$ tends to zero.

 \begin{figure}[t!]
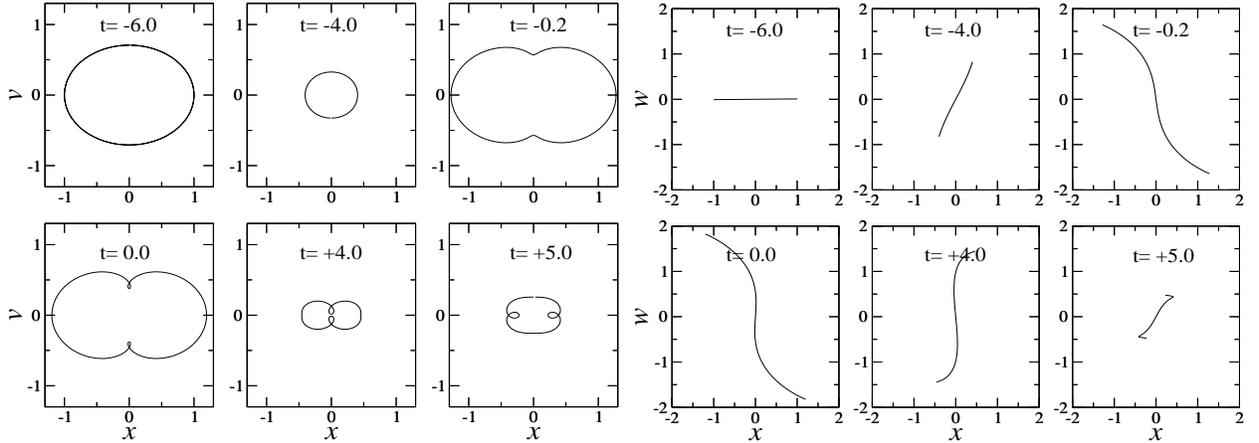

{\centering
\resizebox*{3.2in}{2.3in}{\includegraphics{pos_anti1.eps}}$\,$
\resizebox*{3.2in}{2.3in}{\includegraphics{pos_anti2.eps}}}
\caption{Evolution of the classical analog of
the antisymmetric tensor (\ref{flat_anti}) across the singularity.
The six plots on the left show the behavior in the $xv$
plane, where, inside the Hubble radius, the solution is an oscillating 
elliptical loop. 
Higher modes are excited as $t=0$ is crossed. The plots
on the left show the projection into the $xw$ plane, in
which the solution always appears as a doubled line.
Again, it is clear
from the pictures that higher modes have been excited. For this
case, $t_s=6$ was used. 
}
\label{pos_anti}
\end{figure}

\section{Classical analog of the Antisymmetric Tensor}

To complete our discussion of low mode solutions, we
consider configurations with intermediate angular momentum,
which are the analog of the 
massless antisymmetric tensor states of the quantized string. 
The corresponding flat-space solution was given in 
(\ref{flat_anti}), and its evolution, illustrated in Figure
2, consists of a spinning state whose shape alternates between
an ellipse and a doubled line. The motion is simplest when 
projected onto 
planes spanned by the vectors:
$\hat{x}=(1,0,...,0)$, $\hat{v}=\frac{1}{\sqrt{2}}(0,1,1,0,...,0)$ and
$\hat{w}=\frac{1}{\sqrt{2}}(0,-1,1,0,...,0)$. 
In the $xv$ plane the solution appears as an ellipse which shrinks
to a point and re-expands. 
In the $xw$ plane the motion is similar to that of the 
rotor, but the length of the rotor oscillates.
(see Figure \ref{analogues}).

\begin{figure}[t!]
{\centering
\resizebox*{3.5in}{2.5in}{\includegraphics{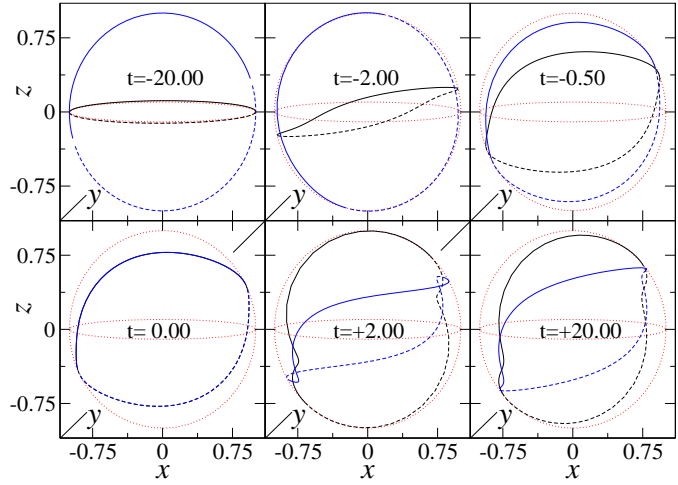}}}
\caption{An incoming classical analog of the antisymmetric tensor state is
represented by left movers (blue lines) and right movers (black lines) 
on the sphere. At $t=0$ the left and right movers coincide in a 
doubled line.
}
\label{sphere_anti}
\end{figure}

If we follow the evolution of such a solution towards 
$t=0$, once it has crossed 
the Hubble horizon, it practically freezes
in size in the $x$ and $w$ axes, and either its position or its
momentum freeze
in the $v$ coordinate. Apart from the special case where the
elliptical motion has a maximum at $t=0$ (which is 
mildly non-analytic, like the case of a circular loop),
the motion is regular across $t=0$ and 
produces a tower of higher excited
modes
(see Figure \ref{pos_anti}). If the
position is the frozen variable in the $v$ direction, there
are fewer excited modes produced, and conversely if the momentum
is the frozen variable.
As 
for the rotor, we can picture the evolution of the antisymmetric
tensor's classical analog in terms of left and right movers on the
sphere (Figure \ref{sphere_anti}).

\vfill\eject 

\section{Quantum evolution across t=0}

In this section we treat the quantum production of excited string
modes in a simple approximation, which may be extended to discuss
more complicated string states. 
As we have already indicated,
for large string states it is reasonable to expect that, provided
the quantum theory remains well-defined, quantum corrections should be
small and our classical calculations should be accurate.

Here we shall limit ourselves to an illustrative example, 
consisting of a truncation of the full theory to the simplest
oscillation mode, namely a circular loop. 
The method we use generalizes to more interesting excited states
and we shall study this in future work.  We start from
the classical equations of motion for a loop,
\be
\label{opeq}
{dp \over dt}=
-{ t^2\,\rho \over H}, \qquad {d \rho \over dt} ={ p\over H}, \qquad
H^2=p^2+ t^2\rho^2.
\ee
We would like to quantize these equations, but the nonlinearity
renders this difficult. We proceed in the following approximation.
The operator $\hat{H}^2
= \hat{p}^2+ t^2\hat{\rho}^2$ is positive definite, 
and so it is not unreasonable to attempt to approximate 
the terms $H$ in the first two equations 
with the $c$-number quantity,
\be
H_c(t)\equiv \sqrt{\langle \hat{H}^2(t) \rangle}.
\label{semia}
\ee
With this replacement, we have only to solve the {\it linear}
quantum system whose Heisenberg equations of motion are  
\be
\label{opeq1}
{d\hat{p} \over dt}=
-{ t^2\,\hat{\rho} \over H_c(t)}, \qquad 
{d \hat{\rho} \over dt} ={ \hat{p}\over H_c(t)}, 
\ee
where $H_c(t)$ is as yet undetermined. 

The general solution to (\ref{opeq}) may be expressed in terms 
of two linearly independent solutions, defined at some 
initial time $t_0$, as follows:
\ba
\hat{\rho}(t) &=& \hat{\rho}_0 f_\rho(t,t_0)+ \hat{p}_0 g_\rho(t,t_0),\cr
\hat{p}(t) &=& \hat{\rho}_0 f_p(t,t_0)+ \hat{p}_0 g_p(t,t_0),
\label{gensol}
\ea
where $f_{\rho}$, $f_p$, $g_{\rho}$ and $g_p$ satisfy the initial
conditions $f_\rho = 1$, $f_p = 0$, $g_\rho = 0$ and $g_p = 1$
at time $t=t_0$. 

We now wish to express the operators $\hat{\rho}_0$ and $\hat{p}_0$
in terms of creation and annihilation operators for the incoming
vacuum state. To do so, we need to identify the relation between
$\rho$ and $p$ and the properly normalized coordinate and momentum of a
harmonic oscillator at large $|t|$. Noticing that the quantity
as
\be
\hat{\mathcal{P}}={\hat{H}^2 \over 2 |t|} = {\hat{p}^2+ t^2\hat{\rho}^2 \over 2|t|},
\label{enq}
\ee
defined in (\ref{newdef}), 
is an action variable at large $|t|$ (see Section \ref{asymst}),
we identify the proper momentum $\hat{\pi} = \hat{p}/\sqrt{|t|}$ and radius
$\hat{R}=\hat{\rho} \sqrt{|t|}$ as the asymptotic coordinate
and momentum of a corresponding harmonic oscillator.
These variables can then be expressed in terms
of creation and annihilation operators, defined at some large 
negative time $t_0$:
\be
\hat{\rho}_0={(a_0+a_0^\dagger)\over \sqrt{2 |t_0|}}, \qquad 
\hat{p}_0=\sqrt{|t_0|\over 2}{(a_0-a_0^\dagger)\over i}.
\label{crann}
\ee
It is then clear how to compute the function $H_c(t)$,
from the formulae above, for any chosen incoming state. In 
particular, in the incoming vacuum state defined by $a_0 |0\rangle_{in}=0$,
we find
\be
H_c^2(t)= {1\over 2} \left({f_p^2\over |t_0|} + |t_0| g_p^2 +
t^2 \left({f_\rho^2\over |t_0|} + |t_0| g_\rho^2 \right)\right).
\label{hccalc}
\ee
It is straightforward to solve equations (\ref{opeq1}) for the
two independent solutions $f$ and $g$, using (\ref{hccalc})
to define $H_c(t)$. This then provides a self-consistent 
semi-classical approximation to the full quantum evolution. 

\begin{figure}[t!]
{\centering
\resizebox*{3.5in}{2.5in}{\includegraphics{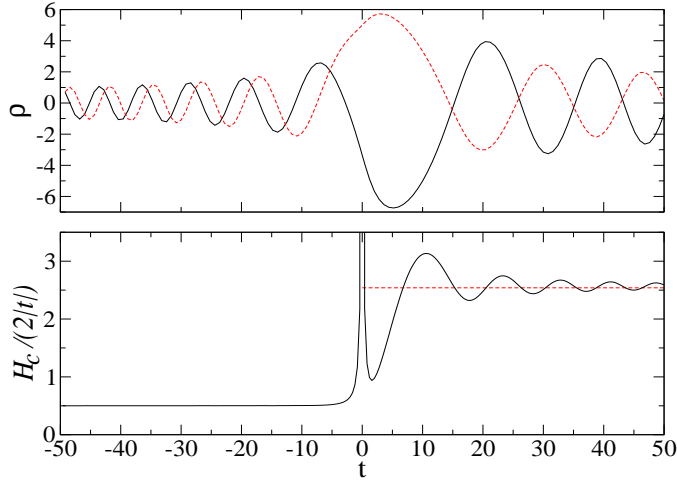}}}
\caption{The quantum creation of higher excitations on a circular loop,
sent in in its ground state. The upper plot shows the 
two independent solutions $f_\rho$ and $g_\rho$ defined in the text.
The lower plot shows the evolution of the `energy'
$H_c/(2|t|)$. This is not well-defined 
at $t=0$, even though the equations of motion are regular there.
For a loop sent in in its vacuum state, with an energy
${1\over 2}$ in these units, the outgoing state has an energy  
of approximately $2.54$.
The increase in energy is due to quantum mode creation
as the loop crosses $t=0$. 
}
\label{qc}
\end{figure}

Figure \ref{qc} shows the results of this calculation, for 
the incoming vacuum state. From formula
(\ref{enq}), we identify $H_c^2/(2|t|)$ as the asymptotic
harmonic oscillator energy at late times, which may in our
chosen units be 
expressed as
$\,_{in}\langle 0|(N_{out}+{1\over 2})|0\rangle_{in}$ where 
$N$ is the usual number operator. Therefore, from the Figure,
we see that in the outgoing state, 
\be
_{in}\langle 0|N_{out}|0\rangle_{in} = |\beta|^2 \approx 2.04,
\ee
where $\beta$ is the usual Bogoliubov coefficient measuring the
amount of particle creation~\cite{birrell}. Similar results
are obtained for excited incoming states. 
One can straightforwardly
compute the Bogoliubov coefficients $\alpha$ and $\beta$ including
their phases, using the same method. The results will be given 
elsewhere.

\section{Summary}

We have herein discussed how classical strings propagate
through the simplest possible big crunch/big bang 
transition in M
theory, and made a start on considering the quantum problem. 
As we have discussed, the usual $\alpha'$ expansion 
must be replaced by an expansion in the string tension,
{\it i.e.} $1/\alpha'$,  once
the Hubble radius falls below the string length. We have
presented an analytic solution of the Nambu equations
as a formal expansion in the string tension,
 which could form the basis for a quantum
treatment of the small time regime.

Using the decomposition into right and left movers, we have
shown
how higher oscillation modes are excited by a finite amount 
as the string crosses singularity. We have 
also given a 
discussion of some special cases where the string
is constrained by symmetry to be static at $t=0$. 
Except for a set of configurations
of measure zero, the classical counterparts of the 
dilaton, graviton and antisymmetric tensor string states
evolve smoothly across $t=0$.

We have made a modest start at describing the quantum evolution
of string across $t=0$, discussing a truncation of the theory
to the lowest mode describing a circular loop. 
The method we have employed appears to be readily
extendible to more complex incoming states, and
to the inclusion of more and more string modes. 
As this is done, we can begin to address the subtle issues
of renormalization which we have so far ignored.
It may also be feasible to go beyond the first 
semiclassical calculation we have described.

Our main conclusion is that classical strings 
pass smoothly through singularities of the simple
type studied here, and the excitations they acquire
in doing so may be readily calculated in a formal
expansion in the string tension. 
While much remains to be done to
develop a full quantum description,
our findings reported here are  
encouraging.

\subsection*{Acknowledgements}
We thank Malcolm Perry, Paul Steinhardt and Andrew Tolley
for discussions. 
This work was supported by CONACYT, SEP and the Cambridge
Overseas Trust (GN) and PPARC (NT). NT is the Darley Professorial
Fellow at Cambridge. 
\appendix
\section{Series expansion around $t=0$}\label{appex}
\subsection{Circular loop}

As explained in Section \ref{lamsec}, in order to study the
motion of the string across $t=0$ we perform a formal
expansion in the string tension. We do so by introducing
a formal parameter $\lambda$ in front of the relevant 
term in the Hamiltonian,  
and then we solve the equations of motion to a given order in
$\lambda$. Finally, we take $\lambda=1$. With this parameter
$\lambda$, equations (\ref{eom_circle_b})-(\ref{eom_circle_c}) take the form 
(\ref{S-circle}), which we repeat here:
\be
{dp \over dt} =
-\lambda { t^2\,\rho\over H},  \qquad {d \rho \over dt} ={ p\over H},
\qquad {d H^2 \over dt} ={2 \over t} (H^2-p^2).
\label{S-circle1}
\ee
From this point, there is only one consistent way to proceed. 
At lowest order in $\lambda$, we solve the above equations
with $\lambda=0$. Then we substitute this solution into
the term involving $\lambda$ and integrate with respect to
$t$ to obtain first $p$, then $H$ and finally $\rho$ to
order $\lambda$. This solution is again substituted into the
right hand side of the first equation, and the procedure
continued. Effectively, we have 
\be \label{eqn1}
{d p^{(n)}\over dt} =-\lambda {t^2\left(\rho\over H\right)^{(n-1)}},\qquad
{ d (t^{-2} (H^2)^{(n)}) 
\over dt}=-2\frac{(p^2)^{(n)}}{t^3},\label{eqn2}
\ee
which can be written as integral equations,
\be \label{eqnP}
p^{(n)}=p_0+\lambda I^{(n-1)},\qquad
(H^2)^{(n)}=p^2_0+\rho_0^2t^2-4\lambda p_0 J^{(n-1)}-2\lambda^2
K^{(n-2)},
\ee
where $p(0)=p_0$, $\rho(0)=\rho_0$ and $I^{(n)}(t),$ $J(^{(n)}t),$ and
$K^{(n)}(t)$ are defined as  
\ba
I^{(n)}(t)=-\int_0^t {\tilde{t}^2\left(\!\rho(\tilde{t})\over
  H(\tilde{t})\!\right)^{(n)}} d\tilde{t}, &&\qquad
J^{(n)}(t)=\int_0^t \frac{I^{(n)}(\tilde{t})}{{\tilde{t}}^3}
d\tilde{t},\\  K^{(n)}(t)=\int_0^t&&
\frac{\left([I(\tilde{t})]^2\right)^{(n)}}{{\tilde{t}}^3} d\tilde{t}, 
\ea
Now equations (\ref{eqnP}) have the desired form.
Finally, we use these solutions for $p$ and $H$ to construct
$\rho$ to the same order in $\lambda$ by expanding the quotient
\be
{d \rho^{(n)}\over dt} 
=\left(\frac{p}{H}\right)^{(n)}\label{eqnR}.
\ee
To zeroth order in $\lambda$, we find
\be
p^{(0)}=p_0,\qquad\qquad H^{(0)}=\sqrt{p^2_0+C^2t^2},
\ee
where $C$ is an integration constant that we will fix at the end,
after setting $\lambda=1$. The solution to $\rho^{(0)}$ is then
given by 
\be
\rho^{(0)}=\rho_0+\frac{p_0}{|C|}\sinh^{-1}\left(\frac{|C|}
{|p_0|}t \right).
\ee
In order to construct the next order solutions, there is a 
change of variables which simplifies the expressions, given by 
\be
\chi\equiv\sinh^{-1}\left(\frac{|C|}{|p_0|}t\right),
\ee
thus the zeroth order solutions look like
\be
p^{(0)}=p_0, \qquad\qquad H^{(0)}=p_0\cosh(\chi), \qquad\qquad
\rho^{(0)}=\rho_0+\frac{p_0}{|C|}\chi. 
\ee
To next order in $\lambda$, we get
\ba
p^{(1)}&=&p_0\Bigg[ 1 + \lambda\frac{p_0\,\rho_0}{4{C}^2}\left( 2\chi
  - \sinh (2\chi) 
  \right) +  \lambda \frac{p_0^2}{4{C}^4}\left( {\chi}^2 + \sinh^2
  (\chi) - \chi\sinh (2\chi) \right)  \Bigg], \nonumber \\ \nonumber
(H^2)^{(1)}&=&p_0^2\Bigg[\cosh^2(\chi)+\lambda\frac{p_0\,\rho_0}{2|C|^3}
  (2\chi\cosh(2\chi)-\sinh(2\chi))\\ && \hspace{1cm} +\lambda\frac{p^2_0}{2C^4}
  \Big(\chi^2\cosh(2\chi)+\sinh^2(\chi)-\chi\sinh(2\chi)\Big)\Bigg]. 
\ea
The parameter $\lambda$ governs the dynamics of solutions near
$t=0$, so it can not appear in the initial conditions. Consequently,
after setting $\lambda=1$ the value for $C$ is determined 
using the equation for the Hamiltonian (\ref{eom_circle_b}),
which implies ${H}''(0)=2\rho_0^2$. From the solution of
$(H^2)^{(0)}$, or equivalently $(H^2)^{(1)}$, at $t=0$ one reads
$C=\rho_0$. Therefore, the solutions for $p^{(1)}$ and $H^{(1)}$ are 
just polynomials in the scale-invariant parameter $|p_0/\rho_0^2|$.

Note that the functions $p^{(1)}$ has an odd and an even piece with
respect to $\chi$. Therefore, the function will be more symmetric
around $t=0$ if $|p_0/\rho_0^2|\gg 1$, and more asymmetric for the
converse value. The odd-$\chi$ part will always have a zero near
$t=0$, which represents either the maximum or the minimum in $\rho$
near $t=0$. The other zero of $p$ depends on the even-$\chi$ piece,
which is close to $t=0$ only for $|p_0/\rho_0^2|\gg 1$. However, the
series is only accurate for small $\chi$ which guarantees $-t^2\rho/
H$ is small after taking $\lambda=1$, so only if the invariant
parameter $|p_0/\rho_0^2|\ll 1$ one can go a little beyond the Hubble
radius crossing. 

By expanding the quotient on the RHS of equation (\ref{eqnR}) to first
order in $\lambda$ we obtain $\rho^{(1)}$, which looks like  
\ba
\rho^{(1)}&=&\rho_0\Bigg[1 + \frac{p_0}{\rho_0\,|C|}\chi + 
  \lambda\frac{p_0^2}{4C_0^4}\left( \chi(2\tanh (\chi)-\chi)-\sinh^2(\chi)
  \right) \\&& \hspace{.7cm}- \lambda\frac{p_0^3}{12\,\rho_0\,|C|^5}
  \left(9\chi+2\chi^3+3\chi\cosh(2\chi)-3\sinh(2\chi) -  
    6\tanh(\chi) -6\chi^2\tanh(\chi)\right)\Bigg].\nonumber
\ea
Again, $\rho^{(1)}$ has piece which is an even function of $\chi$, but now
there are two terms which are odd-$\chi$ functions. It is due to
this extra order term in $p_0/\rho_0^2$ that the $\rho$ solution is more
symmetric if $|p_0/\rho_0^2|\ll 1$ and asymmetric if $|p_0/\rho_0^2|\gg 1$,
in opposition to $p^{(1)}$.\\
\indent In order to obtain better 
precision and more maxima and minima in $\rho$ with this series, one has
to go beyond first order in $\lambda$. However, the integrals become
harder and harder to solve. Just to illustrate the flavor of the next order
solutions, $\rho^{(2)}$ is given by
\ba
\rho^{(2)}&=&\rho_0\cosh^3(\chi)\Bigg\{ \cosh^3(\chi)+ 
\frac{p_0}{\rho_0^2}\left[\chi\cosh^3(\chi) \right] \\ && +
\frac{p_0^2}{16\rho_0^4} \nonumber
\cosh^2(\chi)\bigg[(-1 + 4\chi^2)\cosh(\chi) +\cosh(3\chi)
  -8\chi\sinh(\chi) \bigg] \\&& \nonumber
-\frac{p_0^3}{768\rho_o^6}
\bigg[32\cosh^3 (\chi)\left({\pi}^2 +12 
  \chi^2 - 24\chi\ln(2\cosh(\chi)+12{\mathrm{Li}_2}(-e^{-2\,\chi}))
  \right)  \\&&  \hspace{1.cm} \nonumber + \chi(
  9\cosh(\chi)-4\cosh(3\chi)-12\cosh(5\chi))+(24{\chi}^2-42)
  \sinh (\chi)  \\ && \hspace{1.cm} 
  \nonumber  +( 120{\chi}^2 -27)  
  \sinh (3\chi) + 15\sinh (5\chi)  \bigg]
\\&& \nonumber+
\frac{{{p_0}}^4}{ 12288{{\rho_0}}^7} \bigg[ 3\left(
  347 + 208{\chi}^2 \right) \cosh (5\chi) + 3\cosh (7\chi) -
  1440\chi\sinh (5\chi)  \\      &&\hspace{1.cm} +  
  128 \cosh ^3(\chi)\left( 144\chi^2\ln(2)-96{\chi}^3 +
  5{\chi}^4 +  \nonumber   48(1+3{\chi}^2)\ln (\cosh (\chi))
  \right)  \\ &&\hspace{1.cm} + \nonumber
  6\cosh (\chi)\left(-647-40{\chi}^2 +(473+1248)\cosh(2\chi)\right) 
  \\ && \hspace{1.cm} \nonumber -
  96\chi\left( 2(17+6{\chi}^2)\sinh (\chi) +
  (65+28{\chi}^2)\sinh (3\chi)\right)\\ && 
  \hspace{1.cm}  -96\left( 24{\cosh (\chi)}^3\left( 
  4{\mathrm{Li}_3}(-e^{-2\chi}) +8{\mathrm{Li}_2}(-e^{-2\chi}) +
  3{\zeta}(3) \right)\right) \bigg] \nonumber\\ \nonumber
&&-\frac{p_0^5}{122880\rho_0^9} \bigg[ 4\cosh (\chi)
  \left( 16\cosh^2(\chi)( 40\pi^2+7\pi^4 +
  480\chi^2  + 720\chi^4- 20\chi^6)\right) \\&&
  \hspace{1.cm}-320\chi^3\cosh(\chi) \left( 
  \nonumber -4+13\cosh(4\chi)+57\cosh(2\chi)+
  192\cosh^2(\chi)\ln(2\cosh(\chi))\right)
  \\&& \hspace{1.cm}+15360\cosh^3(\chi)
  \nonumber \left( (2+6\chi^2)\mathrm{Li}_2(-e^{-2\chi})+
  6\chi\mathrm{Li}_3(-e^{-2\chi})+3\mathrm{Li}_4(-e^{-2\chi})
  \right)
  \\ && \hspace{1.cm}+30\sinh(\chi) 
  \nonumber \left(1204+1824\chi^2+320\chi^4+1845\cosh (2\chi)
  +2560\chi^2\cosh(2\chi)\right)\\ && \nonumber \hspace{1.cm}
  +30\sinh(\chi)\left(448\chi^4\cosh(2\chi)+516\cosh(4\chi)+ 
  480\chi^2\cosh(4\chi)+3\cosh(6\chi)\right)
  \bigg]
\Bigg\}
\ea
where we have already set $\lambda=1$ and $C=\rho_0$. In the above
expression, $\zeta(x)$ is the Riemann Zeta function and
$\mathrm{Li}_n(z)$ is the $n$-degree Polylogarithm function, which is
defined as  
~\cite{mathworld} 
\be
\mathrm{Li}_n(z)\equiv\sum_{k=1}^{\infty}\frac{z^k}{k^n},
\ee 
where $z$ belongs to the open unit disk in the complex plane, and is
defined uniquely for $|z|>1$ by analytic continuation.

\subsection{Expansion in the string tension for a 
general string configuration}

In this Appendix we want to construct a formal expansion in the
string tension for an arbitrary string configuration. 
Our starting point is the Hamiltonian corresponding to
the action (\ref{action1}), 
\be
H= \int d \sigma {\cal H}, 
\qquad {\cal H}= \sqrt{{\cal P}^2 + \lambda t^2 (\vec{x}')^2},
\label{genham}
\ee
where in analogy with the circle, we have chosen to work in
units where $\mu_2 \theta_0=1$ but we have inserted a formal
parameter $\lambda$ in front of the term 
involving the string tension. The Hamiltonian equations 
read as follows:
\be
{\partial_t {\cal P} } = \lambda t^2
\partial_\sigma\left( {\partial_\sigma \vec{x} 
\over {\cal H}} \right),\qquad
\partial_t \vec{x}={{\cal P}\over {\cal H}},\qquad 
\partial_t {\cal H}^2 = {2\over t} \left( {\cal H}^2 -{\cal P}^2 \right).
\ee
The second equation can be written in the integral form  
\be\label{eom_s}
\dot{\vec{x}}=\frac{\vec{p}_0+\lambda{\vec{I}}}{{\cal H}}, 
\ee
where $\vec{p}_0=\vec{p}_0(\sigma)$ is the momentum distribution at $t=0$, and 
\be\label{I}
\vec{I}^{(n)}(t,\sigma)=\int_0^td\tilde{t} \, \tilde{t}^2
\partial_\sigma\left( {\partial_\sigma \vec{x}(\tilde{t},\sigma) 
\over {\cal H}(\tilde{t},\sigma)} \right)^{(n)}
\ee
Similarly, ${\cal H}(t,\sigma)$ also has an expansion with respect to
$\lambda$, given by 
\be
({\cal H}^2)^{(n)}=\vec{p}_0^{\,2}+ D^2_0t^2
-4\lambda\vec{p}_0\cdot\vec{J}^{(n-1)}-2\lambda K^{(n-2)},
\ee
where $D_0^2$ is an integration constant that we will fix after taking
$\lambda=1$, and 
\be\label{J}
\vec{J}^{(n)}(t,\sigma)=\int_0^t
\frac{d\tilde{t}}{\tilde{t}^3}\vec{I}^{(n)}(\tilde{t},\sigma),\qquad
K^{(n)}(t,\sigma)=\int_0^t
\frac{d\tilde{t}}{\tilde{t}^3}\left(\big[\vec{I}(\tilde{t},\sigma)\big]
^2\right)^{(n)}
\ee
With these definitions, equation (\ref{eom_s}) becomes
$\partial_t \vec{x}^{(n)}=\left({{\cal P}\over {\cal
    H}}\right)^{(n)}$, and hence the zeroth order equation in
$\lambda$ is 
\be\label{eom_0th}
{\dot{\vec{x}}}^{(0)}=\frac{\vec{p}_0}{\sqrt{\vec{p}_0^{\,2}+ D^2_0t^2}},
\ee
which is easily solved by
\be\label{sol_0th}
\vec{x}^{(0)}=\vec{x}_0+\frac{\vec{p}_0}{|D_0|}
\,\mathrm{sinh}^{-1}\bigg(\frac{|D_0|}{|\vec{p}_0|}t\bigg),
\ee
where $\vec{x}_0$ is the string shape at $t=0$. After setting
$\lambda=1$ the Hamiltonian equation \ref{genham} implies ${\cal
  H^2}''(0)=2(\vec{x}'(0))^2$, which translates into $D_0=|\vec{x}'(0)|$.

Now, to first order in $\lambda$, we find the equation for $\vec{x}$
is given by
\be\label{eom_1st}
\dot{\vec{x}}^{(1)}=\frac{\vec{p}_0}{\sqrt{\vec{p}_0^{\,2}+
  D_0^2t^2}}\bigg[1+\lambda I^{(0)}+\lambda\frac{2t^2\vec{p}_0\cdot
    \vec{J}^{(0)}}{\vec{p}_0^{\,2}+D_0^2t^2}\bigg],
\ee
and with a similar change of variables to that we introduced before for
the circular loop,
$\chi=\sinh^{-1}\left(\frac{|D_0|t}{|\vec{p}_0|}\right)$, the first
order solution reduces to
\ba\label{S-series}
\vec{x}=\vec{x}^{(0)}+\!\frac{\lambda}{|D_0|}\!\int_0^\chi\!
d\vartheta
\vec{I}^{(0)}(\vartheta,\sigma)+\frac{2\lambda\vec{p}_0}{|D_0|\vec{p}_0^2}
\int_0^\chi \!
\frac{\sinh^2(\vartheta)}{\cosh(\vartheta)}\int_0^\vartheta
  \frac{\cosh(\tilde{\vartheta})}{\sinh^3(\tilde{\vartheta})}
  \vec{p}_0\cdot\vec{I}^{(0)}(\tilde{\vartheta},\sigma)d\tilde{\vartheta}
d\vartheta
\ea
These integrals can be reduced to a single integral by 
changing the order of the integrals, provided no 
convergence issues arise. For situations where the parameter 
$|\vec{p}_0|/\vec{x'}_0|\sim 1$, there is no difficulty and the 
the integrals can be exchanged in the
following way 
\ba
\int_0^\chi d\vartheta\vec{I}^{(0)}(\vartheta,\sigma)&=&
\frac{\vec{p}_0^2}{D_0^2}\int_0^\chi  \nonumber
d\tilde{\vartheta} \sinh^2(\tilde{\vartheta})\cosh(\tilde{\vartheta}) 
(\vartheta-\tilde{\vartheta}) \partial_\sigma\left({\partial_\sigma
  \vec{x}(\tilde{\vartheta},\sigma)
  \over {\cal H}(\tilde{\vartheta},\sigma)} \right)^{(0)},
\\ \nonumber
\int_0^\vartheta
\frac{\cosh(\tilde{\vartheta})d\tilde{\vartheta}}{\sinh^3(\tilde
  {\vartheta})}\vec{I}^{(0)}(\tilde{\vartheta} 
,\sigma)&=&\frac{\vec{p}_0^2}{2\,D_0^2}\int_0^\vartheta
d\tilde{\vartheta} \cosh(\tilde{\vartheta}) \bigg[1
  -\frac{\sinh^2(\tilde{\vartheta})}{\sinh^2(\vartheta)}\bigg]
\partial_\sigma\left({\partial_\sigma \vec{x}(\tilde{\vartheta},\sigma)
  \over {\cal H}(\tilde{\vartheta},\sigma)} \right)^{(0)}.
\ea
We have used these last expressions for the example in Figure (\ref{grav-s-exp}).

In principle, the method we have presented may be extended to second
or higher order in the string tension.
However, the integrals become very complicated and only in simple
cases, like the circular loop, one could actually calculate them analytically.

\begin{figure}[t!]
{\centering
\resizebox*{4.5in}{4.5in}{\includegraphics{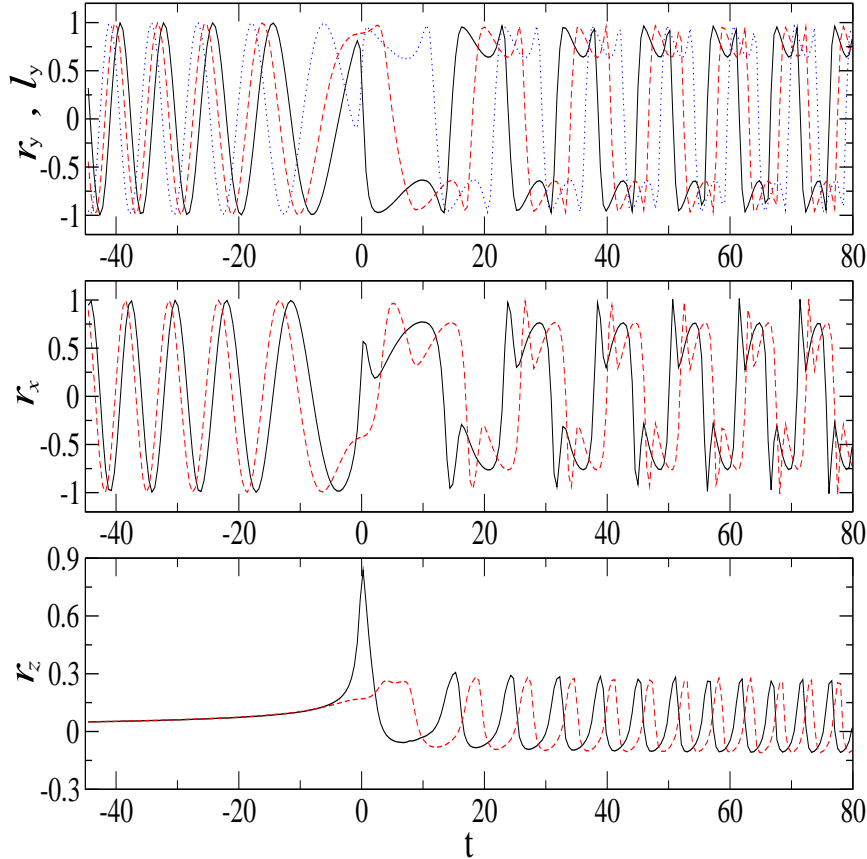}} }
\caption{Evolution of right and left movers for a rotor 
configuration sent in with 
  non-zero transverse velocity. The adiabatic evolution
regime where the 
$\alpha'$ expansion holds is clearly visible: the left and right
movers evolve periodically in proper time. 
 The solid black and the red dashed lines represents a right mover
 at $\sigma=\pi/4$ and at one of the end points ($\sigma=0$)
 respectively. The blue dotted line in the top plot only, describes 
 the left mover at $\sigma=\pi/4$. The third plot shows how the linear
 momentum on the $z$ direction increases and reaches a value close to
 the speed of light for points near the string center 
 (solid black line) and remains almost the same
 for the end point (red dashed line). This is the kink we have
 discussed previously. The same initial parameters as in
 Figure \ref{spheremovers} were used.}
\label{rightmodes}
\end{figure}

\section{Check of numerical evolution}\label{checks}

In order to check our numerical algorithm for following the
string evolution, we have performed a number of tests.
Figure \ref{rightmodes} shows the behavior of right and left
movers for an incoming rotor.  One can readily
distinguish the regime of adiabatic oscillations
from the super-Hubble evolution 
around the singularity. In the former region, the 
right and left movers oscillate as in flat space-time.
Different positions $\sigma$ on
the string are distinguished only 
by a phase difference. Moreover, because
the rotor is isotropic initially, the incoming $x$ and $y$ components
of the right and left movers are equivalent. However, after the
singularity these components behave differently, in agreement with the
loss of isotropy in the $xy$ plane. Furthermore, the spinning doubled line
structure makes the left movers equivalent to
the right movers, with only a phase difference. This structure is
preserved across the singularity, hence only $\alpha_n$, with $n$-odd,
are produced during the transition. Contrary, the $z$ direction gets
modified during the singularity, from constant right and left movers
to a mix of excited states with both, odd and even $n$.
In the $1/\alpha'$ region, the evolution of right and left movers is
completely non-linear and there is not any correlation between right
and left movers, between different components and, perhaps more
interesting, between different points in the string.


\begin{thebibliography}{999}
\bibitem{geom}
\label{geom}
Even if a cosmological constant is present, 
solutions of the form (\ref{milne11d}) still exist in the vicinity
of a brane collision~\cite{turok}.
\bibitem{tension}
More precisely, the membrane tension $\mu_2$ is
related to the eleven dimensional gravitational coupling by a
quantisation condition relating to the four-form flux ~\cite{mukappa}, reading
$\mu_2^3=2\pi^2/(j\kappa_{11}^2)$, with $j$ an integer.
\end{thebibliography}
\end{document}